\documentclass{article}
\usepackage{ijcai26}

\usepackage{times}
\usepackage{soul}
\usepackage{url}
\usepackage[hidelinks]{hyperref}
\usepackage[utf8]{inputenc}
\usepackage[small]{caption}
\usepackage{subcaption}
\usepackage{graphicx}
\usepackage{amsmath}
\usepackage{amsthm}
\usepackage{booktabs}
\usepackage{algorithm}
\usepackage[noend]{algpseudocode}
\usepackage[switch]{lineno}
\usepackage{multirow}
\usepackage{import}
\usepackage{amssymb}
\usepackage{relsize}

\newtheorem{theorem}{Theorem}

\usepackage{siunitx}
\usepackage{booktabs}

\usepackage{amsmath}
\usepackage{tikz,pgffor}
\usetikzlibrary{arrows,backgrounds,calc}
\usetikzlibrary{automata,positioning,arrows,shapes,math,arrows.meta,decorations.pathmorphing}
\tikzstyle{min}=[thick,circle,draw,minimum size=1.3em,inner sep=0em,text centered]
\tikzstyle{state}=[thick,circle,draw,minimum size=1.3em,inner sep=0em,text centered]
\tikzstyle{act}=[thick,circle,draw,fill=black, minimum size=1ex,inner sep=0em,text centered]
\tikzstyle{tr}=[->,thick,rounded corners]

\newcommand{\Nset}{\mathbb{N}}

\newcommand{\Rset}{\mathbb{R}}
\newcommand{\Exp}{\mathbb{E}}
\newcommand{\Var}{\mathrm{Var}}

\newcommand{\RND}{\textrm{RND}}
\newcommand{\RLP}{\textrm{RLP}}
\newcommand{\RSP}{\textrm{RSP}}
\newcommand{\LP}{\textrm{LP}}
\newcommand{\SP}{\textrm{SP}}
\newcommand{\Base}{\textit{Base}}

\newcommand{\Next}{\mathit{next}}
\newcommand{\Update}{\mathit{update}}
\newcommand{\Hit}{\textit{Hit}}
\newcommand{\MHit}{\textit{MHit}}

\newcommand{\Val}{\textit{Val}}
\newcommand{\Run}{\mathit{Run}}

\newcommand{\Mem}{\mathit{Mem}}

\newcommand{\Prob}{\textit{Prob}}

\newcommand{\Init}{\mathit{init}}
\newcommand{\Conf}{\mathit{Conf}}

\newcommand{\Softmax}{\textsc{SoftMax}}
\newcommand{\Act}{\textit{Act}}
\newcommand{\Multihit}{\textsc{AutoHit}}
\newcommand{\Coorhit}{\textsc{CoorHit}}

\newcommand{\En}{\textit{En}}
\newcommand{\Par}{\textit{Par}}

\newcommand{\Dist}{\mathit{Dist}}

\newcommand{\update}{\mathit{update}}

\newcommand{\calF}{\mathcal{F}}

\newcommand{\calL}{\mathcal{L}}

\newcommand{\calR}{\mathcal{R}}

\newcommand{\PTIME}{\mathsf{P}}
\newcommand{\PSPACE}{\mathsf{PSPACE}}

\newcommand{\NP}{\mathsf{NP}}

\newcommand{\suppl}{the Supplementary material}
\usepackage{xspace}
\makeatletter
\DeclareRobustCommand\onedot{\futurelet\@let@token\@onedot}
\def\@onedot{\ifx\@let@token.\else.\null\fi\xspace}

\makeatother

\title{Multiagent Stochastic Shortest Path Problem}

\author{
Martin Jon\'{a}\v{s}
\and
Anton\'{\i}n Ku\v{c}era
\and
Vojt\v{e}ch K\r{u}r
\and
Jan Ma\v{c}\'{a}k
\and
Vojt\v{e}ch \v{R}eh\'{a}k
\affiliations
Faculty of Informatics, Masaryk University, Czechia
\emails
martin.jonas@mail.muni.cz, tony@fi.muni.cz, vojtech.kur@mail.muni.cz, 
macak.jan@mail.muni.cz, rehak@fi.muni.cz
}

\begin{document}

\maketitle

\begin{abstract}
We introduce and study the multiagent stochastic shortest path (MSSP) problem where $k$ agents strive to reach a target state and the task is to minimize the expected time of reaching the target by some agent. We analyze the  computational and strategy complexity of the problem in both autonomous and coordinated settings, and we design efficient strategy synthesis algorithms. The algorithms are experimentally evaluated on instances of increasing size against natural baselines. 
\end{abstract}


\section{Introduction}
\label{sec-intro}

Stochastic Shortest Path (SSP) is a fundamental optimization problem deeply studied in planning and control. The task is to navigate an agent to a target state in a given explicitly represented Markov decision process (MDP) at minimum expected path length (the expected path length can also be interpreted as the expected time to reach a target).
The SSP problem captures a variety of realistic scenarios, including car navigation, drone flying, game playing, etc.

\begin{figure}[t]\centering
    \begin{tikzpicture}[x=7mm, y=5mm, >=stealth', scale=.8,font=\small]
        \node[state] (S)  at (0,0)  {$s$};
        \node[state] (S1) at (4,2 ) {};
        \node[state] (S2) at (4,-2) {};
        \node[state] (S3) at (4,2)  {};
        \node[state] (S4) at (8,-2) {};
        \node[state] (S5) at (8,2)  {$\tau$};
        \node[state] (S6) at (12,0) {};
        \node[act]   (C) at (2,1)  {};
        \node[act]   (A) at (2,-1) {};
        \node[act]   at (6,2) {};
        \node[act]   at (6,-2) {};
        \node[act]   at (10,1)  {};
        \node[act]   at (10,-1) {};
        \node[act]   (B) at (9,3.5) {}; 
        \draw[tr]    (S)  -- node[above, near end] {$1$}  (S1);
        \draw[tr]    (S)  -- node[below, near end] {$0.5$}  (S2);
        \draw[tr]    (A)  -- node[above,near start] {$0.5$}  (S5);
        \draw[tr]    (S2) -- node[above, near end] {$1$}  (S4);
        \draw[tr]    (S3) -- node[above, near end] {$1$}  (S5);
        \draw[tr]    (S4) -- node[below, near end] {$1$}  (S6);
        \draw[tr]    (S6) -- node[above, near end] {$1$}  (S5);
        \draw[tr,-]    (S5) -|  (B);
        \draw[tr]    (B)  -| node[above,near start] {$1$} (S5);
        \node[below of = A, node distance=1.5ex, color=blue] {$b$};
        \node[above of = C, node distance=1.5ex, color=red] {$a$};
    \end{tikzpicture}
    \caption{
    The initial state is $s$, the target is $\tau$, and the numbers denote transition probabilities. If a single agent chooses the action~{\color{red}{$a$}} (or {\color{blue}{$b$}}), the expected time to reach $\tau$ is $2$ (or $2.5$, resp.). Hence, a single agent should choose {\color{red}{$a$}}. Now consider two agents. If both of them choose~{\color{red}{$a$}}, the expected time of reaching $\tau$ by some agent is~$2$. However, if one of them chooses {\color{red}{$a$}} and the other~{\color{blue}{$b$}}, the expectation decreases to $1.5$. (The {\color{blue}{$b$}} action selects between the paths of length $1$ and $4$ uniformly at random. In the first case, the ``blue'' agent arrives to $\tau$ after one step. In the second case, the ``red'' agent arrives to $\tau$ already after two steps. Hence, the expectation is $0.5\cdot 1 + 0.5 \cdot 2 = 1.5$.) Also note that if both agents choose~{\color{blue}{$b$}}, the expectation is $1.75$.}
    \label{fig-two-agents}
\end{figure}
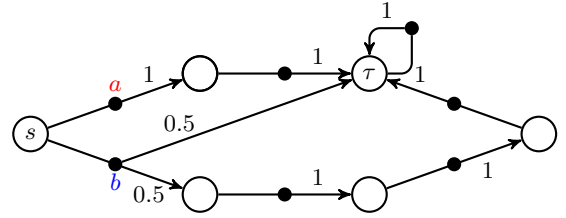

In this work, we initiate the study of \emph{Multiagent Stochastic Shortest Path (MSSP)} problem, where $k$ agents strive to reach a target so that the expected time of reaching the target by some agent is minimized. As in the classical SSP, we consider application scenarios where the MDP is known and represented \emph{explicitly} (hence, there is no need to \emph{learn} the MDP structure). As a simple example, consider the problem of urgent blood transport from a blood bank to a distant hospital by a car driving through a city. The current traffic density determines the probability distribution of the time required to cross individual sections of the road and can be modeled by an MDP whose size is proportional to the number of road sections. Hence, planning an optimal route for a \emph{single} car is an instance of the SSP problem, and an optimal moving strategy~$\sigma$ can be quickly computed by existing algorithms. In order to increase the chance of timely delivery, \emph{multiple cars} (from the same or different blood banks) can be dispatched simultaneously. Clearly, the aim is to minimize the expected time of the \emph{first arrival} (the blood doses delivered by the cars arriving later can be stored for later use). A natural approach to solving this optimization problem is to compute the above strategy $\sigma$ for each car, i.e., to solve the SSP problem independently for each car.  However, this is \emph{not} optimal in general, and in some cases it is even the \emph{worst} possible solution. A simple demonstrative example is given in Fig.~\ref{fig-two-agents}. 

More generally, MSSP models a class of \emph{time-critical delivery problems} where the item can be replicated and sent across multiple paths to improve time performance and reliability. 
The primary research question considered in this work is \emph{``How to compute an optimal solution for a given MSSP instance?''} Apart from inventing efficient algorithms, we analyze the \emph{computational complexity} of MSSP and determine the \emph{type of strategies} required for optimal solutions.

In the above application scenario, the cars can be autonomous or coordinated by a central control room. When the agents are UAVs operating in a hostile environment, the coordination may be unachievable. Therefore, we consider the MSSP problem for both \emph{autonomous} and \emph{coordinated} agents.

Even if agents' coordination is achievable, it can be expensive. Therefore, we also analyze \emph{the benefits of coordination for solving MSSP}, i.e., the ratio between the best solutions achievable in the autonomous/coordinated setting.

\paragraph{Our Contribution.}
We start by analyzing the computational complexity of the MSSP problem and identifying the type of strategies required for optimal solutions.

For the \emph{Coordinated MSSP} problem, we show that an optimal coordinated strategy for $k$~agents in a given MPD $M$ is computable in time \emph{polynomial} in the size of $M$ and \emph{exponential} in~$k$. For every \emph{fixed}~$k$, the problem is solvable in polynomial time, but the degree of the bounding polynomial grows linearly in~$k$. Furthermore, an optimal strategy can be constructed so that it is \emph{memoryless} (the decision depends only on the current positions of the agents) and  \emph{deterministic} (the choice of actions is uniquely determined in every step). 
We also show that the problem is \emph{$\PSPACE$-hard}, and 
hence the exponential blowup in $k$ is \emph{unavoidable}. More precisely, we prove it is $\PSPACE$-hard to decide the existence of a coordinated strategy such that the expected time to reach a target by some agent is bounded by a given constant.

For the \emph{Autonomous MSSP} problem, we show that an optimal autonomous strategy profile for $k$~agents (i.e., a \mbox{$k$-tuple} of independent single-agent strategies) exists, but each agent may require memory to ``remember'' some information about the history of previous moves. 
We also prove that \emph{finite-memory randomized} strategy profiles are sufficient for achieving \mbox{$\varepsilon$-optimal} solutions for an arbitrarily small $\varepsilon>0$. Hence, for practical purposes, it suffices to consider only finite-memory randomized strategy profiles. Then, we show that the problem of deciding the existence of an autonomous strategy profile for $k$ agents such that the expected time to reach a target by some agent is bounded by a given constant is $\NP$-hard \emph{even for $k=2$ and even if the class of admissible profiles is restricted to memoryless profiles}. Hence, the Autonomous MSSP problem is not fixed-parameter tractable in $k$, and it is also \emph{conceptually} harder (in particular, there is no apparent way of computing an $\varepsilon$-optimal profile for a given $\varepsilon \geq 0$). Our algorithmic solution is based on the following:
\smallskip

\noindent
\textbf{(1)}
We show that for every memory size~$m$, the problem of synthesizing the best randomized profile where every strategy in the profile uses at most $m$~memory states is reducible to the problem of computing the best randomized memoryless profile. 
Hence, the algorithmic core of the problem is the synthesis of the best memoryless randomized profile (this problem is still $\NP$-hard even for two agents, see above).
\smallskip

\noindent
\textbf{(2)}
In the next step, we show how to express the expected time to reach a target by some agent as a differentiable function in the parameters representing memoryless randomized profiles. This is the crucial step enabling the application of state-of-the-art techniques of differentiable programming, which is the core of our \Multihit\ algorithm. The efficiency of \Multihit\ is bought by the loss of optimality guarantees (this is unavoidable due to the $\NP$-hardness of the problem).
\smallskip



In the \emph{experimental part} of our paper, we evaluate our synthesis algorithms on instances of increasing size against natural baselines, and we compare the efficiency of solutions obtained in the coordinated and autonomous setting. The instances resemble city maps with a grid-like structure with random delays at certain locations. To avoid bias towards specific MDP classes, the instances are generated randomly.

\paragraph{Related Work.} The SSP problem was introduced in \cite{EZ:SPP-JBE} in the framework of pursuit-evasion games. 
The SSP problem is solvable efficiently by the standard MDP strategy synthesis techniques (i.e., value iteration, policy iteration, and linear programming) under some structural conditions \cite{Puterman:book,BT:SSP-Analysis-MOR,Alfaro:min-max-reachability}. The solution algorithms and their complexity depend also on the properties of the reward function (see, e.g., \cite{BBDGS:SSP-weight-bounded-props}). More recent works study SSP under some additional conditions on the probability distribution of the accumulated reward \cite{PSB:SSP-Variance-Penalized,HK:staying-on-budget,KM:ConditionalMP-Reach-MDP,ADBA:Risk-aware-SPP,MT:Mean-Variance-MDP}. See also \cite{RRS:SSP-VMCAI2015} for an overview. The SSP problem in an \emph{unknown} environment has been studied in various variants, and the solution is mostly constructed by appropriate reinforcement learning techniques. A recent overview of these works is given in \cite{Rosenberg:PhD}. 

Multi-agent problems related to SSP have been studied only recently. In 
\cite{NO:MultiTarget-MDP-TAC}, an algorithm for minimizing the expected time of visiting \emph{all} targets in a given set by a swarm of autonomous agents is proposed. A multi-agent extension of pursuit-evasion games is introduced in \cite{ZPB:PursuitEvasionMultiagent-FRAI}. The problem of minimizing the \emph{total time} of reaching all targets by autonomous agents in an \emph{unknown} environment is studied in \cite{TH:MultiAgentCongestion,ChTH:Regret-MSSP-NeurIPS}. Recent works also study multi-agent variants of infinite-horizon objectives (such as mean-payoff or steady-state objectives) in explictly presented MDPs \cite{JKKM:steady-state-multiagent-IJCAI,DBM-MultiagentSurveillance,KKKMNR:ResilientMultiAgent-IJCAI}. Related objectives such as expected average reward have also been studied for decentralized  POMDPs (see, e.g., \cite{JWXL:DecPOMDP-expected-reward-TAC}).
%
To the best of our knowledge, there are no previous works on MSSP. 


\section{Preliminaries}
\label{sec-prelim}

We assume familiarity with basic notions of probability theory and Markov chain theory. We use $\Nset$ to denote the sets of all non-negative integers and $\Nset_{\infty}$ to denote the set $\Nset \cup \{\infty\}$. For a given finite or countably infinite set~$A$, the set of all 
probability distributions over $A$ is denoted by $\Dist(A)$.
We say that $\mu \in \Dist(A)$ is \emph{Dirac} if $\mu(a) =1$ for some $a \in A$.


\paragraph{Markov Chains.}
A \emph{Markov chain} is a pair $C=(S,A)$ where $S$ is a finite or countably infinite set of \emph{states} and $A : S\times S \to [0,1]$ is a \emph{stochastic matrix} such that $\sum_{t \in S}A(s,t) = 1$ for every $s \in S$.

A \emph{path} in $C$ is a finite or infinite sequence of states. Infinite paths are called \emph{runs}, where a run is also understood as a mapping $\omega : \Nset \to S$. For each finite path $w$, we use $\Run_w$ to denote the set of all runs starting with~$w$. To every $s \in S$ we associate the probability space $(\Run_s,\calF_s,\Prob_s)$ such that $\calF_s$ is the $\sigma$-algebra generated by all $\Run_w$ where $w$ is a finite path initiated in $s$, and $\Prob_s$ is the unique probability measure satisfying $\Prob_s(\Run_w) = \prod_{i=0}^{n-1} A(s_i,s_{i+1})$ for all finite paths $w = s_0,\ldots,s_n$ such that $s_0 = s$.



\paragraph{Markov Decision Processes (MDPs).}
A Markov decision process (MDP) is a triple $M = (S, \Act, P)$ where $S$ and $\Act$ are non-empty finite sets of \emph{states} and \emph{actions}, and
\mbox{$P : S \times \Act \times S \to [0,1]$} is a \emph{transition function}. We require that, for all $s \in S$ and $a \in \Act$, the sum $\sum_{t \in S} P(s,a,t)$ is equal either to~$0$~or~$1$. In the latter case, we say that $a$ is \emph{enabled} in~$s$. The set of all actions enabled in $s$ is denoted by $\En(s)$. We require that $\En(s) \neq \emptyset$ for all $s \in S$.  We say that $t \in S$ is \emph{reachable} from $s \in S$ if there is a finite sequence $s_0,\ldots,s_n$ such that $s_0 = s$, $s_n =t$, and for every $i<n$ there is $a_i \in \Act$ such that $P(s_i,a_i,s_{i+1}) > 0$.

\paragraph{Strategies and Strategy Profiles.} Let $M = (S, \Act, P)$ be an MDP. Consider $k\geq 1$ agents moving among the vertices of $M$ by performing the actions of~$M$. The agents are either \emph{autonomous} or \emph{coordinated}, i.e., each agent acts either independently or coordinates its decisions with the other agents, respectively (the coordination is usually implemented by a central control instructing the agents).

We start by defining a coordinated strategy for $k\geq 1$ agents. Let $\Mem$ be a non-empty set of \emph{memory states} where some information about the history of previous agents' moves is stored. A \emph{configuration} is a tuple $(s_1,\ldots,s_k,m)$ where $s_1,\ldots,s_k \in S$ are the states currently visited by the agents and $m \in \Mem$ is the current memory state. We use $\Conf$ to denote the set of all configurations.  A \emph{coordinated strategy} for $k\geq 1$ agents is a triple $\sigma = (\mu,\Next,\Update)$ where $\mu \in \Mem$ is an \emph{initial memory state} and 
\begin{eqnarray*}
  \Next &:& \Conf \to \Dist(\Act^k),\\ 
  \Update &:& \Conf \times \Act^k \to \Dist(\Mem)
\end{eqnarray*}
are functions defining the next actions taken by the agents and the associated memory update. We require that for all $\gamma \in \Conf$ and $\alpha \in \Act^k$ such that $\Next(\gamma)(\alpha) > 0$, we have that the action $\alpha_i$ is enabled in the underlying state $s_i$ of the configuration $\gamma$ for all $i \in \{1,\ldots,k\}$.

Note that the choice of the next tuple of actions is randomized. The current memory state is updated according to the current configuration and the chosen tuple of actions. We say that $\sigma$ is \emph{deterministic} if $\Next$ and $\Update$ use only Dirac distributions. Furthermore, we say that $\sigma$ is \emph{finite-memory} if the underlying set $\Mem$ is finite, and \emph{memoryless} if $\Mem$ is a singleton. Note that if $\sigma$ is memoryless, then $\Update$ is trivial and $\Next$ can be seen as a function $S^k \to \Dist(\Act^k)$. Slightly abusing our notation, we formally define memoryless strategies as functions $\sigma : S^k \to \Dist(\Act^k)$, i.e., we avoid using the $\Next$ and $\Update$ functions completely.

An \emph{autonomous strategy profile} for $k$~agents is defined as a $k$-tuple $\pi = (\sigma_1,\ldots,\sigma_k)$ where every $\sigma_i$ is a coordinated strategy for \emph{one} agent (see above). Since each $\sigma_i$ controls only the agent~$i$, the behaviour of each agent is completely \emph{autonomous}. We use $\Conf_i = S \times \Mem_i$ to denote the set of all configurations of agent~$i$, where $\Mem_i$ is the set of memory states of $\sigma_i$. We say that $\pi$ is \emph{finite-memory} or \emph{memoryless} if every $\Mem_i$ is finite or a singleton, respectively. 

\paragraph{Random Variable $\MHit$.} 

Let $M = (S, \Act, P)$ be an MDP, $k \geq 1$ the number of agents, and let
$\iota_i \in S$, $T_i \subseteq S$ be the initial/target states 
of agent~$i$ for every $i \in \{1,\ldots,k\}$.

Let $\sigma = (\mu,\Next,\Update)$ be a strategy for $k$~agents, and let   
$\Init = (\iota_1,\ldots,\iota_k,\mu)$ be the \emph{initial configuration}. We define a Markov chain $M_\sigma = (\Conf,A_\sigma)$ where the stochastic matrix $A_\sigma$ is defined as follows: for all $\gamma,\delta \in \Conf$ such that   $\gamma = (s_1,\ldots,s_k,n)$ and $\delta = (t_1,\ldots,t_k,m)$, we put
{\small
\[
   A_\sigma(\gamma,\delta) = \sum_{\alpha \in \Act^k} \Next(\gamma)(\alpha) \cdot \update(\gamma,\alpha)(m) \cdot \prod_{i=1}^k P(s_i,\alpha_i,t_i)\,. 
\]}%

Let $\MHit : \Run_{\Init} \to \Nset_{\infty}$ be a random variable over the runs in $M_\sigma$ initiated in $\Init$ such that $\MHit(\omega)$ is either the least $j \in \Nset$ such that $\omega(j)_i \in T_i$ for some $i \in \{1,\ldots,k\}$, or~$\infty$ if there is no such $j$. We use $\Exp_\sigma[\MHit]$ to denote the expected value of $\MHit$. Hence, if the agents are controlled by~$\sigma$, the expected number of actions executed before some agent visits a target state is equal to $\Exp_\sigma[\MHit]$.

Now consider an autonomous strategy profile $\pi = (\sigma_1,\ldots,\sigma_k)$ for~$k$ agents. For every $i \in \{1,\ldots,k\}$, let $\mu_i$ be the initial memory state of $\sigma_i$ and $\Init_i = (\iota_i,\mu_i)$ the initial configuration of agent~$i$.
%
%
Each $\sigma_i$ determines a Markov chain $M_{\sigma_i}$ and the probability space over $\Run_{\Init_i}$ in the way described above. Let $\Prob_\pi$ be the probability measure in the product probability space over the set $\Run_{\Init_1} \times \cdots \times \Run_{\Init_k}$ of all \emph{multiruns}.
Slightly overloading our notation, we define a random variable $\MHit$ over the multiruns such that $\MHit(\omega_1,\ldots,\omega_k)$ is either the least $j \in \Nset$ such that $\omega_i(j) \in T_i \times \Mem_i$ for some $i \in \{1,\ldots,k\}$, or~$\infty$ if there is no such $j$. We use $\Exp_\pi[\MHit]$ to denote the expected value of $\MHit$. Hence, if the agents move according to the  profile $\pi$, then $\Exp_\pi[\MHit]$ is the expected number of actions executed before some agent visits a target state. 

\paragraph{Multiagent Stochastic Shortest Path (MSSP) Problem.} 
In this work, we study the \emph{Coordinated} and \emph{Autonomous MSSP} problems. In both cases, a problem instance consists~of
\begin{itemize}
  \item an MDP $M = (S, \Act, P)$, 
  \item an integer $k \geq 1$ (the number of agents),  
  \item initial/target states $\iota_i \in S$, $T_i \subseteq S$ such that $\iota_i \not\in T_i \neq \emptyset$ for every $i \in \{1,\ldots,k\}$. 
\end{itemize}

The task is to construct a coordinated strategy $\sigma$ (in the case of Coordinated MSSP) or an autonomous strategy profile $\pi$ (in the case of Autonomous MSSP) for $k$ agents minimizing the expected value of $\MHit$. 

For a given $\varepsilon \geq 0$, we say that $\sigma^*$ and $\pi^*$ are \emph{$\varepsilon$-optimal} if $\Exp_{\sigma^*}[\MHit]$ and $\Exp_{\pi^*}[\MHit]$ are $\varepsilon$-close to $\inf_{\sigma} \Exp_{\sigma}[\MHit]$ and $\inf_{\pi} \Exp_{\pi}[\MHit]$, where $\sigma$ and $\pi$ range over all coordinated strategies and autonomous strategy profiles for $k$ agents, respectively. $0$-optimal $\sigma^*$ and $\pi^*$ are called \emph{optimal}.

\paragraph{Price of Autonomy.}
For every MSSP instance, the \emph{price of autonomy} is defined as 
the ratio of $\inf_{\pi} \Exp_\pi[\MHit]$ to $\inf_{\sigma} \Exp_\sigma[\MHit]$ f. Hence, the price of autonomy measures the benefits of coordination for a given MSSP instance, where higher value means higher benefits.

\section{The Complexity of MSSP}
\label{sec-complex}

In this section, we analyze the computational/strategy complexity of the Coordinated and Autonomous MSSP problems.


\paragraph{Single-Agent Case.}
We begin by recalling known results about the single-agent case, i.e., the SSP problem (see, e.g., \cite{Puterman:book})).

Let $M = (S, \Act, P)$ be an MDP, $k = 1$ the number of agents, $\iota \in S$ an initial state and $T \subseteq S$ a set of target states such that $\iota \not\in T \neq \emptyset$. We have the following:
\begin{itemize}
    \item[(a)] There exists an optimal memoryless deterministic strategy for the agent computable in polynomial time by the LP of Fig.~\ref{fig:lp-single}.
    \item[(b)] For a given memoryless strategy $\sigma$ for one agent, we have that $\Exp_{\sigma}[\MHit]$ is computable in polynomial time by solving the 
    system of linear equations of Fig.~\ref{fig:lin-equations}.
  
\end{itemize} 
More precisely, let $S_{\calR}$ be the maximal $C \subseteq S$ satisfying the following conditions:
\begin{itemize}
    \item if $s \in C$, then there is $t \in T$ reachable from $s$,
    \item if $s \in C\smallsetminus T$, then there is $a \in \En(s)$ such that $t \in C$ for all $t \in S$ where $P(s,a,t)>0$.
\end{itemize}
Furthermore, for every $s \in S_{\calR}$, we use $\En_{\calR}(s)$ to denote the set of all $a \in \En(s)$ such that $P(s,a,t) > 0$ implies $t \in S_{\calR}$.
If $\iota \not\in S_{\calR}$, then the expected time of visiting a target configuration from the initial configuration is infinite for every strategy $\sigma$. Otherwise, let $x_s^*$, $s \in S_{\calR}$ be the solution to the LP of Fig.~\ref{fig:lp-single}, and let $\sigma^*$ be a memoryless deterministic strategy where $\sigma^*(s)(a) = 1$ for an action $a \in \En(s)$ such that
$x_s^* = 1 + \sum_{t \in S_{\calR}} P(s,a,t)\cdot x_t^*$. Then $\sigma^*$ is optimal and $\Exp_{\sigma^*}[\MHit] =  x_{\iota}^*$.

The set $S_{\sigma}$ used in the system of Fig.~\ref{fig:lin-equations} 
is the maximal $C \subseteq S$ such that 
\begin{itemize}
    \item if $s \in C$, then a state of $T$ can be reached from $s$ with positive probability in the Markov chain $M_\sigma$,
    \item if $s \in C {\smallsetminus} T$, $\sigma(s)(a) {>} 0$, and $P(s,a,t) {>} 0$, then $t \in C$. 
\end{itemize}
If $\iota \not\in S_{\sigma}$, then $\Exp_{\sigma}[\MHit] = \infty$.
Otherwise, $\Exp_{\sigma}[\MHit] =  y^*_{\iota}$, where $y^*_s$, $s \in S_{\sigma}$ is the unique solution of the system of linear equations of Fig.~\ref{fig:lin-equations}.


\begin{figure}[t]\small
\[
\begin{array}{llll}
\textbf{maximize}  & \displaystyle\sum\limits_{s \in S_{\calR}} x_s &\\[3ex]
\text{subject to}  & x_s = 0,  &s \in T,\\[1ex]
& \displaystyle x_s \leq 1 + \sum_{t\in S_{\calR}} P(s,a,t)\cdot x_t, & 
\parbox[t]{5em}{%
$s \in S_{\calR} \smallsetminus T$,\\
$a \in \En_{\calR}(s)$}
\end{array}
\]
\caption{A linear program for computing an optimal strategy for a single agent.}
\label{fig:lp-single}
\end{figure}

\begin{figure}[t]\small
\[
\begin{array}{llll}
 y_s = 0 & s \in T,\\[1ex]
 \displaystyle y_s = 1 + \sum_{a \in \En(s)} 
     \left(
        \sigma(s)(a) \cdot \sum_{t \in T} P(s,a,t) \cdot y_t
     \right)
 & s \in S_\sigma \smallsetminus T
\end{array}
\]
\caption{A system of linear equations evaluating a given memoryless strategy $\sigma$ for one agent.}
\label{fig:lin-equations}
\end{figure}

\paragraph{Coordinated MSSP Problem.}

The problem of constructing an optimal coordinated strategy for $k\geq 1$ agents is easily reducible to the SSP problem discussed above. For an MSSP instance $M = (S,\Act,P)$, $k$, $\iota_i$, $T_i$ where $i \in \{1,\ldots,k\}$, we construct an SSP instance consisting of an MDP $M' = (S^k,\Act^k, P')$ where
{\small%
\begin{equation*}
  P'((s_1,\ldots,s_k),(a_1,\ldots,a_k),(t_1,\ldots,t_k))
  = \prod_{i=1}^k P(s_i,a_i,t_i),
\end{equation*}
\normalsize}%
the initial state is $(\iota_1,\ldots,\iota_k)$, and the target states are all $(t_1,\ldots,t_k) \in S^k$ such that $t_i \in T_i$ for some $i \in \{1,\ldots,k\}$.
There is a natural one-to-one correspondence between memoryless coordinated strategies for $k$~agents in $M$ and memoryless strategies for one agent in $M'$, and this correspondence preserves the expected value of~$\MHit$. Hence, an optimal coordinated strategy for $k$ agents in $M$ corresponds to an optimal strategy for one agent in $M'$ computable by the LP of Fig.~\ref{fig:lp-single}. Observe that the size of this LP is polynomial in the size of $M$ and exponential in~$k$. We obtain the following:

\begin{theorem}
  \label{thm-coordinated-upper-bound}
   For every MSSP instance, there exists an optimal memoryless deterministic coordinated strategy computable in time polynomial in the size of the MDP $M$ and exponential in the number of agents~$k$. For every \emph{fixed~$k$} the strategy is computable in polynomial time.
\end{theorem}
In the following, the above described algorithm for computing an optimal memoryless deterministic strategy for a given MSSP instance is referred to as \Coorhit.

Now we show that the exponential blowup in~$k$ is unavoidable (assuming $\PTIME \neq \PSPACE$), because the Coordinated MSSP problem is $\PSPACE$-hard. More precisely, we have the following:

\begin{theorem}
    \label{thm-coordinated-NP-hard}
   Let $M$, $k$, $\iota_i$, $T_i$ where $1 \leq i \leq k$ be an MSSP instance and $B$ a rational bound. The problem of whether there is a coordinated strategy $\sigma$ for~$k$ agents s.t.{} \mbox{$\Exp_\sigma[\MHit] \leq B$} is $\PSPACE$-hard. 
\end{theorem}

A proof of Theorem~\ref{thm-coordinated-NP-hard} is non-trivial and can be found in \suppl.

\paragraph{Autonomous MSSP Problem.}
 
We start by observing that optimal autonomous profiles exist and may require strategies with memory of arbitrarily large size. 

The class of all autonomous strategy profiles is denoted by $\Pi$. Furthermore, for every $n \geq 1$, we use $\Pi_n$ to denote the class of all profiles $\pi = (\sigma_1,\ldots,\sigma_k)$ such that the underlying $\Mem_i$ of every $\sigma_i$ contains at most $n$ memory states. In particular, $\Pi_1$ is the class of all memoryless profiles. Recall that $\pi^*$ is \emph{optimal} if $\Exp_{\pi^*}[\MHit] = \inf_{\pi \in \Pi} \Exp_{\pi}[\MHit]$. Furthermore, we say that 
$\pi^*$ is \mbox{\emph{$\Pi_n$-optimal}} if $\pi^* \in \Pi_n$ and $\Exp_{\pi^*}[\MHit] = \inf_{\pi \in \Pi_n} \Exp_{\pi}[\MHit]$. Note that the existence of an optimal and a \mbox{$\Pi_n$-optimal} profile is not immediately clear. It is established in the next theorem.

\begin{theorem}
\label{thm-optimal}
For every MSSP instance, there exist an optimal profile and a $\Pi_n$-optimal profile for every $n \geq 1$.
\end{theorem}

The existence of an optimal profile is proven by repeatedly selecting converging subsequences of distributions occurring in an infinite sequence of profiles $\pi_1,\pi_2,\ldots$ such that $\lim_{i \to\infty} \Exp_{\pi_i}[\MHit] = \inf_{\pi' \in \Pi} \Exp_{\pi'}[\MHit]$. A similar technique is used to prove the existence of a $\Pi_n$-optimal profile. The  details are in \suppl.

Now we show that optimal autonomous profiles may require strategies with memory of arbitrarily large size. However, for every $\varepsilon > 0$, there exists an $\varepsilon$-optimal profile with finite memory.
\begin{theorem}
\label{thm-memory}
   For every $n \geq 1$, there exist an instance of the Automonous MSSP problem with two agents and a profile $\pi \in \Pi_{n+1}$ such that
   \(
       \Exp_\pi[\MHit]  <  \inf_{\pi' \in \Pi_n} \Exp_{\pi'}[\MHit]
   \). Furthermore, for every instance of the Automonous MSSP problem with $k$~agents and every $\varepsilon > 0$ there exists an $\varepsilon$-optimal finite-memory profile.
\end{theorem}
Furthermore, in \suppl\ we show that randomized profiles are \emph{strictly more powerful} than deterministic profiles when the agents use memory of a given size. 

Now we analyze the computational complexity of Autonomous MSSP. 
The next theorem says that the problem of computing a profile $\pi \in \Pi_n$ for an MDP $M$ minimizing $\Exp_\pi[\MHit]$ can be efficiently reduced to the problem of computing a \emph{memoryless} profile $\bar{\pi}$ minimizing $\Exp_{\bar{\pi}}[\MHit]$ in an effectively constructible MDP $\bar{M}$. 

\begin{theorem}
\label{thm-reduce}
   Let $M = (S,\Act,P)$ be an MDP, $k \geq 1$, and $\Mem$ a set of memory states of size $n \geq 1$. Then there is an MDP $\bar{M} = (S{\times} \Mem, \Act {\times} \Mem, \bar{P})$ such that
   \begin{itemize}
    \item for every profile $\pi = (\sigma_1,\ldots,\sigma_k)$ for $M$ where every $\sigma_i$ is a strategy with memory $\Mem$, there exists a memoryless profile $\bar{\pi} = (\bar{\sigma}_1,\ldots,\bar{\sigma}_k)$ for $\bar{M}$ such that the Markov chains $M_{\sigma_i}$ and $\bar{M}_{\bar{\sigma}_i}$ are identical for every $i \in \{1,\ldots, k\}$,
    \item for every memoryless profile $\bar{\pi} = (\bar{\sigma}_1,\ldots,\bar{\sigma}_k)$ for $\bar{M}$, there exists a profile $\pi = (\sigma_1,\ldots,\sigma_k)$ for $M$ where every $\sigma_i$ is a strategy with memory $\Mem$ such that the Markov chains $\bar{M}_{\bar{\sigma}_i}$ and $M_{\sigma_i}$ are identical for every $i \in \{1,\ldots, k\}$.
   \end{itemize} 
\end{theorem}
Intuitively, the MDP $\bar{M}$ is obtained from $M$ by encoding the elements of $\Mem$ into the states of $\bar{M}$. Hence, instead of synthesizing a finite-memory profile for $M$, we may synthesize a memoryless profile $\bar{\pi}$ for $\bar{M}$ and then ``translate'' $\bar{\pi}$ into a finite-memory profile for~$M$ so that
 $\Exp_{\bar{\pi}}[\MHit]$ in $\bar{M}$ is equal to $\Exp_{\pi}[\MHit]$ in $M$. In other words, the algorithmic core of the MSSP problem is the \emph{synthesis of memoryless profiles}. Unfortunately, the synthesis problem is $\NP$-hard even for two agents and memoryless profiles. 
 
\begin{theorem}
\label{thm-hardness-autonomous}
Let $M$ be an MDP and  $\iota_i$, $T_i$ where $i \in \{1,2\}$ initial/target states for two agents. Let $B$ be a rational bound. The problem of whether there exists an autonomous strategy profile $\pi$ for two agents such that $\Exp_\pi[\MHit] \leq B$ is $\NP$-hard. The problem is $\NP$-hard even if the class of admissible profiles is restricted to memoryless profiles.
\end{theorem}
A proof is obtained by a tricky reduction from a variant of the Boolean satisfiability problem, see \suppl.
Finally, we show that the price of autonomy can be arbitrarily large, even for two agents.

\begin{theorem}
\label{thm-price}
   For every $B \in \Nset$, there exists an MSSP instance with two agents such that the price of autonomy is at least~$B$.
\end{theorem}

\section{Minimizing $\Exp_\pi[\MHit]$}
\label{sec-sol}

The results of the previous section show that the problem of computing an optimal coordinated strategy for a given MSSP instance is fixed-parameter tractable in the number of agents~$k$. However, there is no straightforward algorithmic solution for the Autonomous MSSP. In this section, we address this challenge by  
designing \Multihit, an efficient profile synthesis algorithm for the Autonomous MSSP.  We begin by summarizing the principle limitations. 

Due to Theorems~\ref{thm-memory} and~\ref{thm-reduce}, we can safely restrict our attention to the synthesis of \emph{memoryless} profiles. Since synthesizing optimal memoryless profiles is $\NP$-hard even for two agents (see Theorem~\ref{thm-hardness-autonomous}), it cannot be done efficiently unless $\PTIME = \NP$. Hence, \emph{efficiency inevitably leads to losing optimality guarantees}. Furthermore, the synthesis should not be limited to deterministic profiles because randomized profiles are strictly more powerful (see the remarks after Theorem~\ref{thm-memory}). On the other hand, the synthesis may take advantage of the explicit MDP representation by analyzing and utilizing relevant structural properties of~$M$. The design of \Multihit\ reflects these observations. 

We start by designing an efficient \emph{evaluation procedure} computing $\Exp_\pi[\MHit]$ for a \emph{given} memoryless profile~$\pi$. 
Note that a naive evaluation based on constructing a product Markov chain and solving the linear system of Fig.~\ref{fig:lin-equations} is exponential in~$k$. A more efficient procedure is given below. 


For the rest of this section, we fix an MSSP instance consisting of an MDP $M = (S,\Act,P)$, $k \geq 1$, and $\iota_i \in S$, $T_i \subseteq S$ for every $i \in \{1,\ldots,k\}$.


\paragraph{Evaluating Memoryless Profiles.}


Let $\pi = (\sigma_1,\ldots,\sigma_k)$ be a memoryless profile. Recall that each $\sigma_i$ is formally a function $S \to \Dist(\Act)$ and determines a Markov chain $M_{\sigma_i} = (S,A_i)$.
For the sake of clarity, we use $\Hit_i$ to denote a random variable over the runs of $M_{\sigma_i}$ such that $\Hit_i(\omega)$ is either the least $j \in \Nset$ such that $\omega(j) \in T_i$, or $\infty$ if there is no such~$j$.
Since the agents are independent, we have that
{\small
\begin{equation}
\Exp_\pi[\MHit] \ = \ \sum_{\ell=1}^{\infty} \Prob_\pi[\MHit {\geq} \ell] \ = \ 
\sum_{\ell=1}^{\infty} \prod_{i=1}^k \Prob_{\iota_i}[\Hit_i {\geq} \ell]
\end{equation}}%

For every $i\in \{1,\ldots, k\}$, we fix a target state $\tau_i \in T_i$ and define a transition matrix $\bar{A}_i$ obtained from $A_i$ by changing $\tau_i$ into a sink and modifying transitions ending in a state of $T_i$ so that they end in $\tau_i$. More precisely, we put 
\begin{itemize}
   \item $\bar{A}_i(s,t) = A_i(s,t)$ for all $s,t \not\in T_i$,
   \item $\bar{A}_i(s,\tau_i) = \sum_{t \in T_i}  A_i(s,t)$ for all $s \not\in T_i$,
   \item $\bar{A}_i(s,t) = 0$ for all $s \not\in T_i$ and $t \in T_i {\smallsetminus} \{\tau_i\}$,
   \item $\bar{A}_i(t,t) = 1$, and $\bar{A}_i(t,s) = 0$ for all $t \in T_i$ and $s \neq t$.
\end{itemize}
Then, for every $\ell \geq 0$, we have that $\Prob_{\iota_i}[\Hit_i {\leq} \ell]  =  \bar{A}_i^{\ell}(\iota_i,\tau_i)$ where $\bar{A}_i^0$ is the identity matrix. Hence, 
\begin{equation}
   \Prob_{\iota_i}[\Hit_i {\geq} \ell]  =  1- \bar{A}_i^{\ell-1}(\iota_i,\tau_i)
\end{equation}
for all $\ell \geq 1$, and thus we obtain
\begin{equation}
\Exp_\pi[\MHit] \ = \ 
\sum_{\ell=1}^{\infty} \prod_{i=1}^k \big(1- \bar{A}_i^{\ell-1}(\iota_i,\tau_i)\big)
\label{eq-EMHit-eval}
\end{equation}

Since~\eqref{eq-EMHit-eval} is an infinite sum, it is not algorithmically workable. Therefore, we also consider a truncated version of~\eqref{eq-EMHit-eval} where the range of $\ell$ is restricted to  $\{1,\ldots,\gamma\}$ for a suitable constant $\gamma \in \Nset$. Formally, we define
\begin{equation}
\Exp\langle\gamma\rangle_\pi[\MHit] \ = \ 
\sum_{\ell=1}^{\gamma} \prod_{i=1}^k \big(1- \bar{A}_i^{\ell-1}(\iota_i,\tau_i)\big)
\label{eq-EMHit-eval-gamma}
\end{equation}
Note that $\Exp\langle\gamma\rangle_\pi[\MHit]$ is computable in time \emph{polynomial} in $k$, $\gamma$, and the size of $M$. Furthermore, we show that for every $\varepsilon > 0$, there is efficiently computable $\gamma_\varepsilon$ such that $\Exp_\pi[\MHit] - \Exp\langle\gamma_\varepsilon\rangle_\pi[\MHit] \leq \varepsilon$. Thus, the value of $\Exp_\pi[\MHit]$ can be efficiently  \mbox{$\varepsilon$-approximated} by evaluating the right-hand side of~\eqref{eq-EMHit-eval-gamma}. 

The constant $\gamma_\varepsilon$ is computed as follows. Recall that for every $i \in \{1,\ldots,k\}$, the value of $\Exp_{\sigma_i}[\Hit_i]$ is computable in polynomial time by solving the system of linear equations of Fig.~\ref{fig:lin-equations}. For a given $\varrho \in \Nset$, we put 
\begin{equation*}
\delta\langle\varrho\rangle_i = \Exp_{\sigma_i}[\Hit_i] - \sum_{\ell=1}^{\varrho} (1 - \bar{A}^{\ell-1}_i(\iota_i,\tau_i)) \,.
\end{equation*}
Furthermore, let $X_i = \{1,\ldots,k\} \smallsetminus \{i\}$. Then 
\begin{equation}
   \Exp_\pi[\MHit] - \Exp\langle\varrho\rangle_\pi[\MHit] \ \leq \ \delta\langle\varrho\rangle_i \cdot \prod_{j\in X_i} (1 - \bar{A}^{\varrho}_j(\iota_j,\tau_j))
\label{eq-epsilon}
\end{equation}
Note that~\eqref{eq-epsilon} follows immediately from definitions and the fact that $\bar{A}^{\ell}_j(\iota_j,\tau_j) \geq \bar{A}^{\varrho}_j(\iota_j,\tau_j)$ for all $\ell \geq \varrho$.
Hence, $\gamma_\varepsilon$ 
can be set to the least $\varrho$ such that the right-hand side of~\eqref{eq-epsilon} is bounded by $\varepsilon$ for some $i\in\{1,\ldots,k\}$. Since the right-hand side of~\eqref{eq-epsilon} decreases exponentially in~$\varrho$, the value of $\gamma_\varepsilon$ is small for most instances.






\paragraph{Representing Memoryless Profiles.}
Now we show how to represent memoryless strategies by vectors of real-valued parameters so that $\Exp\langle\gamma\rangle_\pi[\MHit]$ becomes a differentiable function of these parameters.

Let $i \in \{1,\ldots, k\}$ be an agent index. For all $s \in S$ and $a \in \En(s)$, we fix a fresh parameter (variable) $X_{i,s,a}$ ranging over $\Rset$. We use $\Par_i$ to denote the vector of all $X_{i,s,a}$. Furthermore, for every $s \in S$, we use $\Par_{i,s}$ to denote the vector of all $X_{i,s,a}$ where $a \in \En(s)$.

Let $\Softmax$ be the standard softmax function transforming a vector of real values into a probability distribution, i.e., a vector of the same dimension over $(0,1]$ whose sum is equal to one. 
The values of $\Par_i$ represent a memoryless strategy $\pmb{\sigma}_i$ such that $\pmb{\sigma}_i(s)(a) = \Softmax(\Par_{i,s})(a)$ for every $a \in \En(s)$. In other words, for every $s \in S$, the values of $\Par_{i,s}$ are transformed into a probability distribution over $\En(s)$ using $\Softmax$. Note that $\pmb{\sigma}_i$ can be seen as a function of $\Par_i$. We also use $\pmb{\bar{A}}_i$ and $\pmb{\bar{A}}_i^\ell$ to denote the matrices $\bar{A}_i$ and $\bar{A}_i^\ell$ determined by $\pmb{\sigma}_i$ (see the previous paragraphs). Again, $\pmb{\bar{A}}_i$ and $\pmb{\bar{A}}_i^\ell$ are seen as functions of $\Par_i$. More concretely, for all $s,t \in S$, the element $\pmb{\bar{A}}_i^\ell(s,t)$ is a function of $\Par_i$ involving $\Softmax$, multiplication, addition, and some constants. Note that $\pmb{\bar{A}}_i^\ell(s,t)$ is differentiable.


Let $\Par_1,\ldots,\Par_k$ be parameters whose values represent a memoryless profile $\pmb{\pi} = (\pmb{\sigma}_1,\ldots,\pmb{\sigma}_k)$ in the way described above. We use $\pmb{\Exp\langle\gamma\rangle_\pi[\MHit]}$ to denote the value
of $\Exp\langle\gamma\rangle_\pi[\MHit]$ for the profile $\pmb{\pi}$. Note that $\pmb{\Exp\langle\gamma\rangle_\pi[\MHit]}$ is a differentiable function of $\Par_1,\ldots,\Par_k$.

\algnewcommand{\Inputs}[1]{%
  \State \textbf{Inputs:}
  \Statex \hspace*{\algorithmicindent}\parbox[t]{.8\linewidth}{\raggedright #1}
}
\algnewcommand{\Outputs}[1]{%
  \State \textbf{Outputs:}
  \Statex \hspace*{\algorithmicindent}\parbox[t]{.8\linewidth}{\raggedright #1}
}
\algnewcommand{\Hyper}[1]{%
  \State \textbf{Numerical hyperparameters:}
  \Statex \hspace*{\algorithmicindent}\parbox[t]{.8\linewidth}{\raggedright #1}
}
\algnewcommand{\Initialize}[1]{%
\State \textbf{Initialize:}
\Statex \hspace*{\algorithmicindent}\parbox[t]{.8\linewidth}{\raggedright #1}
}
\begin{algorithm}[t]
	\small
	\caption{The \Multihit\ Synthesis Algorithm}
	\label{alg:multihit}
	\begin{algorithmic}[1]
        \Inputs{MDP $M = (S,\Act,P)$,\\
                the number of agents $k \geq 1$,\\
                $\iota_i \in S$, $T_i \subseteq S$ for all $i \in \{1,\ldots,k\}$
        } 
        \Outputs{A memoryless profile $\pi$ for $M$ and $k$ agents\\
                 \textit{Val} such that  $\Exp_\pi[\MHit] - \varepsilon \leq \textit{Val} \leq \Exp_\pi[\MHit]$}
        \Hyper{$\textit{Steps},\varepsilon,\gamma$}
       \State $\Par_1,\ldots,\Par_k \gets \textsc{InitParams}(M,k)$\label{line:init}
	    \ForAll{$\mathit{index} \in \{1,\ldots, \textit{Steps}\}$}\label{loop:start}
            \State $\textit{grad} \gets \nabla \,\pmb{\Exp\langle\gamma\rangle_\pi[\MHit]}(\Par_1,\ldots,\Par_k)$\label{line:grad} 
		      \State $\Par_1, \ldots,\Par_k \gets \textsc{Optimize}(\textit{grad},\Par_1,\ldots,\Par_k)$\label{loop:end}
		 \EndFor
      \State $\textit{Val} \gets \textsc{Evaluate}(\pmb{\pi},\varepsilon)$\label{line:eval}
		\State \Return $\pmb{\pi}$, \textit{Val}
	\end{algorithmic}
\end{algorithm}

\paragraph{\Multihit\ Algorithm.} 
The \Multihit\ algorithm (Algorithm~\ref{alg:multihit}) is based on minimizing 
$\pmb{\Exp\langle\gamma\rangle_\pi[\MHit]}$ by gradient descent from a suitable initial profile.

More precisely, \Multihit\ starts by invoking an abstract function
$\textsc{InitParams}(M,k)$ initializing the parameters $\Par_1,\ldots,\Par_k$ to
values representing an initial profile $\pi_0$ (line~\ref{line:init}). In our
experiments, $\pi_0$ is either a \emph{random profile} or a \emph{precomputed profile} (in the latter case, $\pi_0$ is computed by some efficient method based on analyzing the structure of~$M$, see Section~\ref{sec-experiments}). For the random profile $\pi_0$, the value of each parameter $X_{i,s,a}$ is sampled from the normal distribution with expected value~$0$ and variance~$1$. 



The main optimization loop at lines \ref{loop:start}--\ref{loop:end} is terminated after $\textit{Steps}$ iterations, where $\textit{Steps}$ is a hyperparameter. In each step, the gradient of $\pmb{\Exp\langle\gamma\rangle_\pi[\MHit]}$ at $\Par_1,\ldots,\Par_k$ is computed (line~\ref{line:grad}) and then passed to a gradient-based optimizer that computes modified parameter values (line~\ref{loop:end}). Our implementation uses \textsc{PyTorch} library \cite{PyTorch} to compute the gradient, and $\textsc{Optimize}(\textit{grad},\Par_1,\ldots,\Par_k)$
is implemented by \textsc{Adam} optimizer \cite{adam}.
At line~\ref{line:eval}, the value of $\Exp_\pi[\MHit]$ for the resulting profile $\pmb{\pi}$ is computed up to the precision $\varepsilon$ by the function 
$\textsc{Evaluate}(\pmb{\pi},\varepsilon)$ whose implementation is described in the paragraph ``Evaluating Memoryless Profiles'' above.

\section{Experiments}
\label{sec-experiments}

Our experimental evaluation aims to answer the following fundamental research
questions: \emph{What is the quality of strategies and strategy profiles synthesized by our \Coorhit\ and \Multihit\ algorithms? What is the efficiency/scalability of these algorithms?} 
We also aim to compare the quality of solutions obtained in the coordinated/autonomous setting (i.e., estimate the price of autonomy).


\paragraph{Benchmarks.} We designed a scalable family of
benchmarks resembling cities with congestion at some crossroads. The
benchmarks, parameterized by their length $l$ and a ``congestion factor''
$p_c$, consist of states $s_{x, y}$ for all $1 {\leq} x {\leq} l$, $1 {\leq} y {\leq} 5$ and actions $\mathit{left}$, $\mathit{right}$, $\mathit{up}$, $\mathit{down}$. Each state is \emph{congested} with probability $p_c$ (to avoid bias towards specific instances, the subset of congested states is determined randomly). In
non-congested states, the actions work as expected, moving the agent left,
right, up, and down, respectively. In congested states, the actions work as
expected only with probability $p \in \left[\frac{1}{8}, \frac{1}{2}\right]$ and
with probability $1{-}p$ do not move the agent (i.e., they lead back to the same
state). Thus, congested states model random delays at road segments. The goal is to reach the target state $s_{l, 3}$ from $s_{1,3}$.

\paragraph{Setup.} We used a Linux machine with AMD Ryzen 7 PRO 5750G CPU and 32
GB of RAM. All experiments were executed without using GPUs. Each execution has
wall clock time limit of $5$~minutes. In all experiments, the hyperparameters are set to
$\mathit{Steps} = 1000$ and $\varepsilon = 10^{-9}$. The hyperparameter~$\gamma$ is
set to the number of states of the MDP. This ensures that each simple path
from the initial state to the target state is reflected in the objective
function\footnote{We also experimented with other values of $\gamma$, the
  results are available in \suppl.} $\pmb{\Exp\langle\gamma\rangle_\pi[\MHit]}$.

\paragraph{Results for \Multihit.} For each $l \in \{ 10, 20, 30, 40, 50 \}$, we generated $10$ random benchmarks with $p_c {=} 0.2$. Thus, we considered $50$
benchmarks in total. As a baseline, we use the profile $\pi_{\mathrm{LP}}$
consisting of the \emph{optimal single-agent strategies} obtained by solving
the LP of Fig.~\ref{fig:lp-single}.
We ran \Multihit\ for every $k \in \{ 1, 5, 10, 15, 20 \}$. All
\Multihit~executions took less than $85$ seconds of wall time, even for $20$ agents and grids
with $l {=} 50$ containing $250$ states and $890$ actions.  

We executed \Multihit~with two initial parameter settings representing a random profile $\pi_{\RND}$ and a precomputed profile $\pi_{\RLP}$ which is a ``randomized version'' of $\pi_{\LP}$. Note that since $\pi_{\LP}$ is deterministic, it cannot be precisely translated into the parameter values due to the use of \Softmax. Instead, each parameter $X_{i,s,a}$ representing $\pi_{\RLP}$ is sampled from the normal distribution with the expected value equal to $10$ or  $0$, depending on whether the strategy $\sigma_i$ in the profile $\pi_{\LP}$ selects the action $a$ in $s$ or not, respectively (the variance is~$1$). For each MSSP instance, we 
generate~$5$ randomly sampled versions of $\pi_{\RND}$ and $\pi_{\RLP}$, and we 
use $\Val_{\RND}$ and $\Val_{\RLP}$ to denote the \emph{average} $\Val$ returned by $\Multihit$ for these $5$ versions of $\pi_{\RND}$ and $\pi_{\RLP}$, respectively. We compare $\Val_{\RND}$ and $\Val_{\RLP}$ against the baseline $\Base = \textsc{Evaluate}(\pi_{\mathrm{LP}}, \varepsilon)$.


\begin{figure}[t]
  \centering

  \includegraphics[width=\columnwidth]{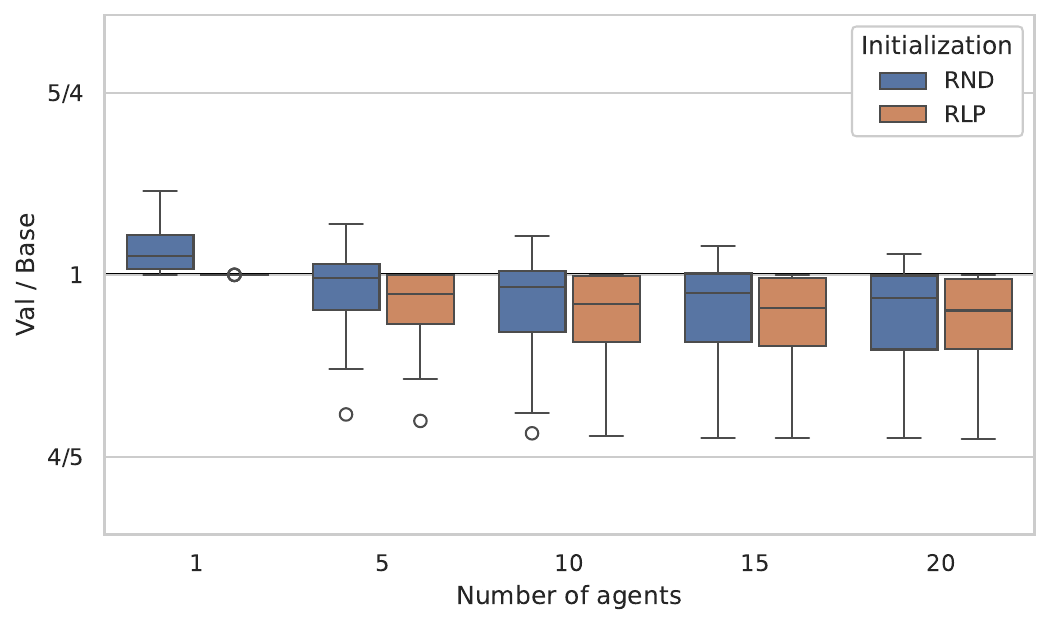}
  
  \caption{We report the $\Val_{\RND}/\Base$ (blue) and $\Val_{\RLP}/\Base$ (brown) ratios. The points below the line  $y=1$ correspond to benchmarks where the ratio is smaller than~$1$, i.e., the profiles computed by \Multihit\ outperform the baseline.}
  \label{fig:results_gd_all}
\end{figure}

The results are presented in Fig.~\ref{fig:results_gd_all}. The value $\Val_{\RND}$ is on average 97.5~\% of $\Base$ and is even 81.9~\% in some cases. 
For $\Val_{\RLP}$, this improves to 96.0~\% on
average and 81.8~\% in the best case. Hence, \Multihit\ computes
strategy profiles that \emph{outperform} the baseline, decreasing the expected value of $\MHit$ by almost 20~\% in some cases. When initialized with $\pi_{\RLP}$, the profile computed by \Multihit\ outperforms the baseline \emph{in the vast majority of the benchmarks}.


We executed similar experiments also for another precomputed initial profile $\pi_{\SP}$ consisting of strategies following a graph-theoretic shortest path in the MDP. However, the quality of the resulting profiles computed by \Multihit\ tends to be worse than for $\pi_{\RND}$ and $\pi_{\RLP}$. We refer to \suppl\ for details.


\begin{figure}[t]
  \centering

  \includegraphics[width=\columnwidth]{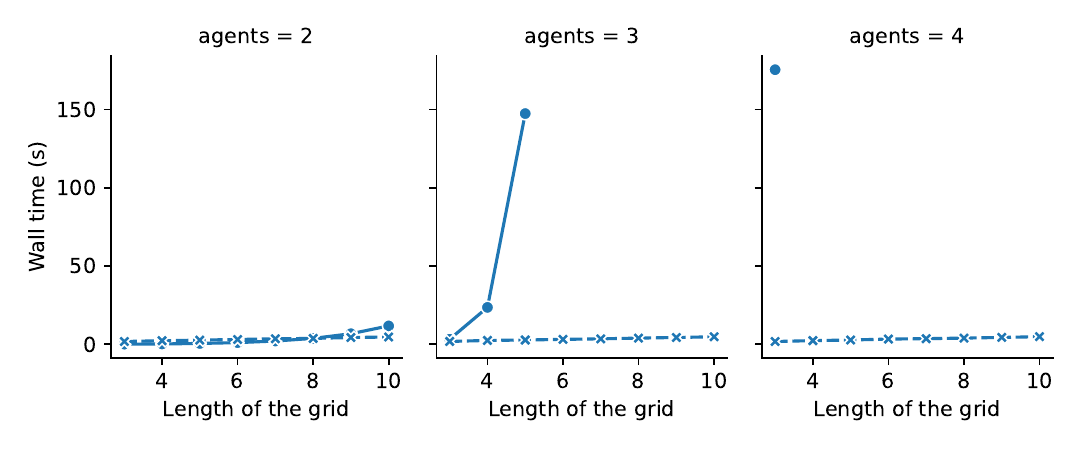}
  
  \caption{Average execution times of \Coorhit\ (solid line) and \mbox{\Multihit} (dashed line).}
  \label{fig:results_coordinated}
\end{figure}

\paragraph{Results for \Coorhit.} Our \Coorhit\ implementation is based on Gurobi \cite{gurobi} LP~solver. Since \Coorhit\ does not scale so well in the number of agents, we use smaller grids.We generated $10$ random benchmarks for each $l$
between $3$ and $10$. To ensure that there are some congested crossroads on the
shortest path, we used $p_c = 1$. We ran the algorithms with $2$, $3$, and $4$ agents.

The average runtimes of \Coorhit\ and \Multihit\ (initialized with $\pi_{\RLP}$ profile) are shown in
Fig.~\ref{fig:results_coordinated}. Note that \Multihit~scales
significantly better than \Coorhit, consistently with our theoretical results. For $4$ agents,  \Coorhit\ did not terminate for any grid with
$l > 3$ within $5$ minutes. Hence, \Multihit~is significantly more efficient than \Coorhit.


In the cases where both algorithms terminated, the expected values of $\MHit$ for the resulting strategies and strategy profiles differ by $0.01$ on average and at most by $0.08$. This shows that the price of autonomy is rather low in the considered benchmarks. 


\section{Conclusions}
Our initial study of the MSSP problem reveals the fundamental complexity barriers for designing efficient algorithmic solutions in both coordinated and autonomous setting. Our \mbox{\Multihit} algorithm successfully overcomes the barriers in the (more challenging) autonomous setting and produces high-quality solutions. 

For our randomly generated benchmarks, \mbox{\Multihit} \emph{consistently} improves the expected value of $\MHit$ over the baseline $\Exp_{\pi_{\LP}}[\MHit]$
where $\pi_{\LP}$ is the profile consisting of optimal single-agent strategies. In some cases, this improvement is \emph{significant}. This shows that despite the high computational complexity of the MSSP problem, there are efficient synthesis algorithms producing solutions outperforming the natural baseline $\Exp_{\pi_{\LP}}[\MHit]$.

The questions whether a similarly efficient synthesis algorithm exists also for the coordinated setting and whether alternative approaches can lead to even better synthesis algorithms certainly deserve an adequate research effort left for future work.  
\vfill


\bibliographystyle{named}
\bibliography{str-long,concur}

\newpage
\appendix

\begin{center}
   \huge\bf Supplementary Material
\end{center}

\section{Proof of Theorem~2}

In this proof, we first show that a certain variant of the quantified Boolean formula problem (QBF) is PSPACE-hard. We then prove PSPACE-hardess of the Coordinated MSSP problem by reduction from that QBF variant. A truth assignment is a mapping from the set of all propositional variables $Var = \{x_0, x_1, x_2, \ldots\}$ to the set of truth values $\{0,1\}$. Every truth assignment is associated with a unique extended truth assignment, which is a mapping from the set of all (quantified) Boolean formulae to the set of truth values $\{0,1\}$. In this proof, we do not distinguish between a truth assignment and its corresponding extended truth assignment (we use the same symbol for both). Let $\odot$ denote the ternary logical connective whose semantics is the indication of whether exactly one of the three parameters is true (e.g., $\odot(x_1, x_2, x_3)$ evaluates to $1$ under a truth assignment $\nu'$ such that $\nu'(x_1)=\nu'(x_2)=0$ and $\nu'(x_3)=1$, but evaluates to $0$ under a truth assignment $\nu''$ such that $\nu''(x_1)=\nu''(x_2)=\nu''(x_3)=1$). The aforementioned QBF variant used in this proof restricts the input formulae to be of the form $Q_1x_1Q_2x_2\ldots Q_kx_k\ \varphi$ where each $Q_i$ stands either for $\exists$ or $\forall$ and $\varphi$ is a conjunction of subformulae of the form $\odot(x_{i_1}, x_{i_2}, x_{i_3})$ (all propositional variables occurring in $\varphi$ are quantified). In the next two paragraphs, we prove PSPACE-hardness of this QBF variant by showing a simple polynomial-time reduction from another QBF variant where the input formulae are restricted to be of the form $Q_1x_1Q_2x_2\ldots Q_{\ell}x_{\ell}\ \psi$ for some Boolean formula $\psi$ in 3-CNF (i.e., each clause of $\psi$ has exactly three distinct literals). This QBF variant is known to be PSPACE-complete.\footnote{See \emph{Oijid, N. (2025). Bounded Degree QBF and Positional Games. In: Finocchi, I., Georgiadis, L. (eds) Algorithms and Complexity. CIAC 2025. Lecture Notes in Computer Science, vol 15680. Springer, Cham. https://doi.org/10.1007/978-3-031-92935-9\_8}}

Let $\Psi\equiv Q_1x_1Q_2x_2\ldots Q_{\ell}x_{\ell}\ \psi$ be an arbitrary input formula of the latter QBF variant. Without restrictions, we assume that $\psi$ contains no tautological clauses. We show how to efficiently transform $\Psi$ to an equivalent formula $\Phi$ of the desired form. For each clause of $\psi$ of the form $(x_{i_1}\lor x_{i_2}\lor x_{i_3})$ (i.e., without negations), replace this clause by 
\begin{eqnarray*}
&& \odot(x,y,w)\land\odot(y,z,w)\land\odot(z,x,w)\land\odot(x_{i_1},x,t_{i_1})\\
&\land & \odot(x_{i_2},x,t_{i_2})\land\odot(x_{i_3},x,t_{i_3})\land\odot(t_{i_1},r,d)\\ 
&\land& \odot(t_{i_2},u,e) \land \odot(t_{i_3},v,f)\land\odot(d,e,f)
\end{eqnarray*}
where $t_{i_1}$, $t_{i_2}$, $t_{i_3}$, $w$, $x$, $y$, $z$, $r$, $u$, $v$, $d$, $e$, $f$ are fresh propositional variables, and append 
\[
\exists t_{i_1} \ \exists t_{i_2}\ \exists t_{i_3}\ \exists w\ \exists x\ \exists y \ \exists z \ \exists r \ \exists u \ \exists v \ \exists d \ \exists e \ \exists f
\]
to the quantifier prefix of the formula. Proceed analogously also for clauses containing negative literals $\neg x_i$ (for one or more $i\in\{i_1, i_2, i_3\}$), but use $x_i$ instead of the last occurrence of $t_i$ in the subformula replacing the original clause: for example, the clause $(x_4\lor \neg x_5\lor \neg x_6)$ is replaced by 
\begin{eqnarray*}
&& \odot(x,y,w)\land\odot(y,z,w)\land\odot(z,x,w)\land\odot(x_4,x,t_4)\\
&\land & \odot(x_5,x,t_5)\land\odot(x_6,x,t_6)\land\odot(t_4,r,d)\land\odot(x_5,u,e)\\
&\land& \odot(x_6,v,f)\land\odot(d,e,f)
\end{eqnarray*}

Obviously, the transformation described in the previous paragraph can be implemented by a polynomial-time algorithm (the length of the formula increases by at most a constant multiplicative factor) and the resulting formula $\Phi$ is of the desired form. Let $\nu: Var \to \{0,1\}$ be an arbitrary truth assignment, let $\xi$ denote the formula 
\begin{eqnarray*}
&& \odot(x,y,w)\land\odot(y,z,w)\land\odot(z,x,w)\land\odot(x_{i_1},x,t_{i_1})\\
&\land&\odot(x_{i_2},x,t_{i_2})\land\odot(x_{i_3},x,t_{i_3})\land\odot(t_{i_1},r,d)\\ 
&\land& \odot(t_{i_2},u,e) \land \odot(t_{i_3},v,f)\land\odot(d,e,f)
\end{eqnarray*}
and let $\xi'$ denote the formula 
\[
\exists t_{i_1} \ \exists t_{i_2} \ \exists t_{i_3} \ \exists w \ \exists x \ \exists y \ \exists z \ \exists r \ \exists u \ \exists v \ \exists d \ \exists e \ \exists f\ \xi\,.
\]
In order to show that the described transformation is equivalence-preserving, we are to prove that 
\[ 
\nu(x_{i_1}\lor x_{i_2}\lor x_{i_3}) = \nu(\xi')
\]
(the proof for the cases with negative literals is analogous). Let us first consider the case when $\nu(x_{i_1}\lor x_{i_2}\lor x_{i_3})=0$, which implies that $\nu(x_{i_1})=\nu(x_{i_2})=\nu(x_{i_3})=0$. Assume for the sake of contradiction that $\nu(\xi')=1$. It follows that there exists a truth assignment $\mu$ such that $\mu(x_{i_1})=\nu(x_{i_1})$, $\mu(x_{i_2})=\nu(x_{i_2})$, $\mu(x_{i_3})=\nu(x_{i_3})$ and $\mu(\xi)=1$: let us fix any such $\mu$. It is easy to observe that necessarily $\mu(w)=1$, $\mu(x)=\mu(y)=\mu(z)=0$ and thus $\mu(t_{i_1})=\mu(t_{i_2})=\mu(t_{i_3})=1$, implying $\mu(d)=\mu(e)=\mu(f)=0$, which is in contradiction with $\mu(\odot(d,e,f))=1$. Let us now consider the case when $\nu(x_{i_1}\lor x_{i_2}\lor x_{i_3})=1$. Without loss of generality, we assume that $\nu(x_{i_1})=1$ (the proof in the remaining two cases is analogous). We are to prove $\nu(\xi')=1$. Let $\mu'$ be the variable assignment such that 
\begin{itemize}
  \item $\mu'(w)=1$, 
  \item $\mu'(x)=\mu'(y)=\mu'(z)=0$, 
  \item $\mu'(t_{i_1})=1-\nu(x_{i_1})=0$, 
  \item $\mu'(t_{i_2})=1-\nu(x_{i_2})$, 
  \item $\mu'(t_{i_3})=1-\nu(x_{i_3})$, 
  \item $\mu'(r)=0$, 
  \item $\mu'(d)=1$, 
  \item $\mu'(e)=\mu'(f)=0$, 
  \item $\mu'(u)=\nu(x_{i_2})$, 
  \item $\mu'(v)=\nu(x_{i_3})$,
  \item $\mu'(x')=\nu(x')$ for all remaining variables $x'$.
\end{itemize}
It is easy to verify that $\mu'(\xi)=1$ and thus $\nu(\xi')=1$, which was to be proved.

For the rest of this proof, let us fix an input formula $\Phi\equiv Q_1x_1Q_2x_2\ldots Q_kx_k\ \varphi$, where $\varphi\equiv (C_1\land C_2\land\ldots\land C_n)$ and each $C_j$ is a subformula of the form $\odot(x_{i_1}, x_{i_2}, x_{i_3})$ (these subformulae are often understood as sets of three propositional variables in this proof). We describe a construction (implementable by a polynomial-time algorithm) of an instance of the MSSP problem for $k$ agents and a rational bound $B$ such that a coordinated strategy $\sigma$ achieving $\Exp_\sigma[\MHit] \leq B$ exists if and only if the formula $\Phi$ is true. Since there always exists an optimal coordinated strategy which is deterministic and memoryless (see the main body of the paper), we consider only memoryless strategies in this proof and we understand the term configuration simply as a $k$-tuple of states throughout the rest of this proof (completely omitting the memory state from the tuple).

Our construction is based on an observation of a simple MSSP instance where three agents can choose between two actions enabled in their (shared) initial state. One choice causes the agent to reach the target state after exactly $2$~steps (with probability~$1$). The other choice causes the agent to reach the target state after $1$ step or after $4$ steps with probability $0.5$ for both eventualities. Optimal strategies are precisely those where (with probability $1$) exactly one of the agents makes the first choice, whereas the remaining two agents make the second choice. Technical description of the whole construction follows.

Let $M = (S, \Act, P)$ be an MDP where
\begin{eqnarray*}
S & = & \{\tau\} \cup \{s'_{i,\ell} \mid 1 \leq i \leq k, 1\leq \ell \leq i\}\\
& \cup & \{s'_{i,\top,\ell} \mid 1 \leq i \leq k, i+1 \leq \ell \leq k+1\}\\
& \cup & \{s'_{i,\bot,\ell} \mid 1 \leq i \leq k, i+1 \leq \ell \leq k+1\}\\
& \cup & \{s''_{i,\top,j} \mid 1 \leq i \leq k, 1 \leq j \leq 4n\}\\
& \cup & \{s''_{i,\bot,j} \mid 1 \leq i \leq k, 1 \leq j \leq 4n\},\\[1ex]
\Act  & = &\{\top, \bot\}
\end{eqnarray*}
and the transition function $P$ is defined as follows: 
\begin{itemize}
    \item $P(\tau,\bot,\tau) = 1$; 
    \item $P(s'_{i,\ell},\bot,s'_{i,\ell +1}) = 1$ for all $1 \leq i \leq k$ and $1 \leq \ell \leq i-1$; 
    \item $P(s'_{i,i},\bot,s'_{i,\top,i+1}) = P(s'_{i,i},\bot,s'_{i,\bot,i+1}) = \frac{1}{2}$ for all $1 \leq i \leq k$ such that $Q_i\equiv \forall$; 
    \item $P(s'_{i,i},\top,s'_{i,\top,i+1}) = P(s'_{i,i},\bot,s'_{i,\bot,i+1}) = 1$ for all $1 \leq i \leq k$ such that $Q_i\equiv \exists$; 
    \item $P(s'_{i,\top,\ell},\bot,s'_{i,\top,\ell +1}) = P(s'_{i,\bot,\ell},\bot,s'_{i,\bot,\ell +1}) = 1$ for all $1 \leq i \leq k$ and $i+1 \leq \ell \leq k$; 
    \item $P(s'_{i,\top,k+1},\bot,s''_{i,\top,1}) = P(s'_{i,\bot,k+1},\bot,s''_{i,\bot,1}) = 1$ for all $1 \leq i \leq k$; 
    \item $P(s''_{i,\top,4n},\bot,\tau) = P(s''_{i,\bot,4n},\bot,\tau) = 1$ for all $1 \leq i \leq k$; 
    \item $P(s''_{i,\top,4j-3},\bot,s''_{i,\top,4j-2})=P(s''_{i,\top,4j-2},\bot,s''_{i,\top,4j-1})=P(s''_{i,\top,4j-1},\bot,s''_{i,\top,4j})=P(s''_{i,\bot,4j-3},\bot,s''_{i,\bot,4j-2})=P(s''_{i,\bot,4j-2},\bot,s''_{i,\bot,4j-1})=P(s''_{i,\bot,4j-1},\bot,s''_{i,\bot,4j})=1$ for all $1 \leq i \leq k$ and $1 \leq j \leq n$ such that $x_i \notin C_j$; 
    \item $P(s''_{i,\top,4j},\bot,\tau)=P(s''_{i,\bot,4j},\bot,\tau)=\frac{99}{100}$, 
    \item $P(s''_{i,\top,4j},\bot,s''_{i,\top,4j+1})=P(s''_{i,\bot,4j},\bot,s''_{i,\bot,4j+1})=\frac{1}{100}$ for all $1 \leq i \leq k$ and $1 \leq j \leq n-1$ such that $x_i \notin C_j$; 
    \item $P(s''_{i,\top,4j-3},\bot,s''_{i,\top,4j-2})=P(s''_{i,\top,4j-1},\bot,s''_{i,\top,4j})=P(s''_{i,\bot,4j-2},\bot,s''_{i,\bot,4j-1})=P(s''_{i,\bot,4j-1},\bot,s''_{i,\bot,4j})=1$, 
    \item $P(s''_{i,\top,4j-2},\bot,s''_{i,\top,4j-1})=\frac{1}{100}$, 
    \item $P(s''_{i,\top,4j-2},\bot,\tau)=\frac{99}{100}$, 
    \item $P(s''_{i,\bot,4j-3},\bot,\tau)=\frac{99}{200}$, 
    \item $P(s''_{i,\bot,4j-3},\bot,s''_{i,\bot,4j-2})=\frac{101}{200}$ for all $1 \leq i \leq k$ and $1 \leq j \leq n$ such that $x_i \in C_j$; 
    \item $P(s''_{i,\top,4j},\bot,s''_{i,\top,4j+1})=1$, 
    \item $P(s''_{i,\bot,4j},\bot,s''_{i,\bot,4j+1})=\frac{2}{101}$, 
    \item $P(s''_{i,\bot,4j},\bot,\tau)=\frac{99}{101}$ for all $1 \leq i \leq k$ and $1 \leq j \leq n-1$ such that $x_i \in C_j$; 
    \item the function $P$ maps all the remaining triples to $0$.
\end{itemize}

For each $i \in \{1,\ldots,k\}$, let $T_i = \{\tau\}$ be the set of target states and let $s'_{i,1}$ be the initial state of agent $i$. Furthermore, let 
\[
  B = k + 1 + 1.2601255\cdot \frac{10^{2nk}-1}{10^{2nk}-10^{2(n-1)k}}\,.
\]  

Observe that in the constructed MSSP instance, there is only the $\bot$ action enabled in every state of the MDP $M$ except for states $s'_{i,i}$ such that $Q_i$ is an existential quantifier, where both actions $\top$ and $\bot$ are enabled. After $k$ steps from the beginning, agent $i$ can be either in state $s'_{i,\top,k+1}$ or in state $s'_{i,\bot,k+1}$. Which of these two possibilities will occur is decided in the $i$-th step, when the state of agent $i$ changes from $s'_{i,i}$ to either $s'_{i,\top,i+1}$ or $s'_{i,\bot,i+1}$. If $Q_i$ is an existential quantifier, then the strategy can decide (by choosing either the action $\top$ or $\bot$ for agent $i$) in the $i$-th step whether agent $i$ will be in state $s'_{i,\top,k+1}$ or $s'_{i,\bot,k+1}$ after $k$ steps from the beginning. If $Q_i$ is a universal quantifier, then such decision in the $i$-th step is made randomly (with probability $\frac{1}{2}$ for both possibilities). Once every agent has made $k$ steps from the beginning, the $\top$ action will not be enabled anymore. It follows that after $k$ steps, the strategy starts to be completely irrelevant to the behaviour of the system and each agent starts to be fully independent (as in the Autonomous MSSP).

After $k$ steps from the beginning, the sequence of states visited by agent $i$ must be either of the form $s'_{i,\top,k+1}, s''_{i,\top, 1}, s''_{i,\top, 2}, \ldots, s''_{i,\top, d}, \tau, \tau, \ldots$ or of the form $s'_{i,\bot,k+1}, s''_{i,\bot, 1}, s''_{i,\bot, 2}, \ldots, s''_{i,\bot, d}, \tau, \tau, \ldots$ for some $d\in\{1,2,\ldots,4n\}$. Let $j\in\{1,\ldots,n\}$. Let us first assume $j\neq n$. The MDP $M$ is designed in a way that whenever an agent visits the state $s''_{i,\top,4j-3}$ (or $s''_{i,\bot,4j-3}$), then with probability $0.01$ the agent reaches the state $s''_{i,\top,4j+1}=s''_{i,\top,4(j+1)-3}$ (or $s''_{i,\bot,4j+1}=s''_{i,\bot,4(j+1)-3}$, respectively) after exactly four next steps and with probability $0.99$ the agent instead reaches the target state $\tau$ in at most four next steps: more specifically, if $x_i\notin C_j$, then the agent visiting the state $s''_{i,\top,4j-3}$ (or $s''_{i,\bot,4j-3}$) reaches the target state $\tau$ after exactly four next steps with probability $0.99$, whereas if $x_i\in C_j$, then the agent visiting the state $s''_{i,\top,4j-3}$ (or $s''_{i,\bot,4j-3}$) reaches the target $\tau$ after exactly two steps with probability $0.99$ (or after exactly one step with probability $0.495$ and after exactly four steps with probability $0.495$, respectively). Let us now assume $j=n$. The situation in the case $j=n$ is similar, but instead of reaching $s''_{i,\top,4(j+1)-3}$ or $s''_{i,\bot,4(j+1)-3}$ (which are not states of the MDP $M$) from $s''_{i,\top,4j-3}$ or $s''_{i,\bot,4j-3}$ in exactly four steps, the agent reaches the target state $\tau$ in exactly four steps (the probability of the latter event is thus increased by $0.01$).

Let $\mu: Var \to \{0,1\}$ be an arbitrary truth assignment. Let $\bar{\mu}$ denote the configuration in which the state visited by each agent $i$ is equal to $s'_{i,\top,k+1}$ if $\mu(x_i)=1$ or equal to $s'_{i,\bot,k+1}$ if $\mu(x_i)=0$ (this configuration can occur after exactly $k$ steps from the beginning in the constructed MSSP instance). Let random variable $\MHit_{\mu}$ be the number of steps taken until the target $\tau$ is first reached by some agent from the configuration $\bar{\mu}$. Similarly, let random variable $\Hit_{\mu}^i$ be the number of steps taken until the target $\tau$ is first reached by agent $i$ from the configuration $\bar{\mu}$ (for all $i\in\{1,\ldots,k\}$). Let us inspect the expected value of $\MHit_{\mu}$, denoted by $\Exp[\MHit_{\mu}]$.

\begin{eqnarray*}
\Exp[\MHit_{\mu}] & = & \sum_{\ell=0}^{\infty} \ell \cdot\Prob[\MHit_{\mu} {=} \ell] 
 =  \sum_{\ell=0}^{\infty} \Prob[\MHit_{\mu} {>} \ell]\\
& = & \Prob[\MHit_{\mu} {>} 0]+\sum_{\ell=1}^{4n} \Prob[\MHit_{\mu} {>} \ell]\\
& = & 1+\sum_{j=1}^{n}\Prob[\MHit_{\mu} {>} 4j-3]\\
  &&  +  \sum_{j=1}^{n}\Prob[\MHit_{\mu} {>} 4j-2]\\
  &&  +  \sum_{j=1}^{n}\Prob[\MHit_{\mu} {>} 4j-1]\\
  &&  +  \sum_{j=1}^{n}\Prob[\MHit_{\mu} {>} 4j].
\end{eqnarray*}

Using independence, we get that $\Prob[\MHit_{\mu} {>} \ell] = \prod_{i=1}^k \Prob[\Hit_{\mu}^i {>} \ell]$ for all $\ell\in\Nset$. The previously described properties allow us to derive the following:
\begin{eqnarray*}
\Prob[\Hit_{\mu}^i {>} 4j-3] & = & \Prob[\Hit_{\mu}^i {>} 4j-2]\\
& = & \Prob[\Hit_{\mu}^i {>} 4j-1]\\
& = & \Prob[\Hit_{\mu}^i {>} 4j]\\
& = & 0.01^{j-1}
\end{eqnarray*}
for all $1 \leq i \leq k$ and $1 \leq j \leq n$ such that $x_i \notin C_j$.
Furthermore,
\begin{eqnarray*}
 \Prob[\Hit_{\mu}^i {>} 4j-3] & = & \Prob[\Hit_{\mu}^i {>} 4j-2]\\
  &=&  0.01^{j-1},\\
 \Prob[\Hit_{\mu}^i {>} 4j-1] & = & \Prob[\Hit_{\mu}^i {>} 4j]\\
 &=& 0.01^j = 0.01^{j-1}\cdot 0.01
\end{eqnarray*}
for all $1 \leq i \leq k$ and $1 \leq j \leq n$ such that $x_i \in C_j$ and $\mu(x_i)=1$.
Finally,
\begin{eqnarray*}
 \Prob[\Hit_{\mu}^i {>} 4j-3] & = & 0.01^{j-1},\\
 \Prob[\Hit_{\mu}^i {>} 4j-2] & = & \Prob[\Hit_{\mu}^i {>} 4j-1]\\
 & = & \Prob[\Hit_{\mu}^i {>} 4j]\\
 & = & 0.01^{j-1}\cdot 0.505
\end{eqnarray*}
 for all $1 \leq i \leq k$ and $1 \leq j \leq n$ such that $x_i \in C_j$ and $\mu(x_i)=0$.

In the following part of this proof, we show that $\Exp[\MHit_{\mu}]\geq B-k$, where $\Exp[\MHit_{\mu}]= B-k$ if and only if $\mu(\varphi)=1$. Let us first assume that $\mu(\varphi)=1$. Recall that for every $j\in\{1,2,\ldots,n\}$ there are exactly three values of $i$ such that $x_i \in C_j$, where for exactly one such $i$ we have that $\mu(x_i)=1$. Hence, we obtain 
\begin{eqnarray*}
 \Prob[\MHit_{\mu} {>} 4j-3] & = & \prod_{i=1}^k \Prob[\Hit_{\mu}^i {>} 4j-3]\\
 &=& 0.01^{(j-1)k},\\
 \Prob[\MHit_{\mu} {>} 4j-2] & = & \prod_{i=1}^k \Prob[\Hit_{\mu}^i {>} 4j-2]\\
 &=& 0.01^{(j-1)k}\cdot 0.505^2,\\
\Prob[\MHit_{\mu} {>} 4j-1] &=& \Prob[\MHit_{\mu} {>} 4j]\\
&=& \prod_{i=1}^k \Prob[\Hit_{\mu}^i {>} 4j-1]\\
&=& 0.01^{(j-1)k}\cdot 0.01\cdot 0.505^2\,
\end{eqnarray*}
for every $j\in\{1,2,\ldots,n\}$. We can now infer the following:

\begin{eqnarray*}
 \Exp[\MHit_{\mu}] & = & 1+\sum_{j=1}^{n}\Prob[\MHit_{\mu} {>} 4j-3]\\
   && +\sum_{j=1}^{n}\Prob[\MHit_{\mu} {>} 4j-2]\\
   && +\sum_{j=1}^{n}\Prob[\MHit_{\mu} {>} 4j-1]\\
   && +\sum_{j=1}^{n}\Prob[\MHit_{\mu} {>} 4j]\\
  &=& 1+\sum_{j=1}^{n}0.01^{(j-1)k}(1+0.505^2+\\
  &&0.01\cdot 0.505^2 + 0.01\cdot 0.505^2)\\
  &=& 1+1.2601255\sum_{j=1}^{n}0.01^{(j-1)k}\\
  &=& 1+1.2601255\cdot \frac{10^{2nk}-1}{10^{2nk}-10^{2(n-1)k}}\\
  &=& B-k\,.
\end{eqnarray*}

Let us now assume that $\mu(\varphi)=0$. Our aim is to show that $\Exp[\MHit_{\mu}] > B-k$. We first show that 
\begin{eqnarray*}
 \Exp[\MHit_{\mu}] &=& 1+\sum_{j=1}^{n}\Prob[\MHit_{\mu} {>} 4j-3]\\
 && +\sum_{j=1}^{n}\Prob[\MHit_{\mu} {>} 4j-2]\\
 && +\sum_{j=1}^{n}\Prob[\MHit_{\mu} {>} 4j-1]\\
 && +\sum_{j=1}^{n}\Prob[\MHit_{\mu} {>} 4j]\\
  &\geq& 1+\sum_{j=1}^{n}1.2601255\cdot 0.01^{(j-1)k} = B-k
\end{eqnarray*}
by comparison of the respective summands. Let $j\in\{1,2,\ldots,n\}$ be an arbitrary index of summation. The sum
\begin{eqnarray*}
&& \Prob[\MHit_{\mu} {>} 4j-3] + \Prob[\MHit_{\mu} {>} 4j-2]\\
&+&\Prob[\MHit_{\mu} {>} 4j-1] + \Prob[\MHit_{\mu} {>} 4j]\\
\end{eqnarray*}
is equal to
\begin{eqnarray*}
&& \prod_{i=1}^k \Prob[\Hit_{\mu}^i {>} 4j-3]
+  \prod_{i=1}^k \Prob[\Hit_{\mu}^i {>} 4j-2]\\
&+&  \prod_{i=1}^k \Prob[\Hit_{\mu}^i {>} 4j-1]
+ \prod_{i=1}^k \Prob[\Hit_{\mu}^i {>} 4j]\,.
\end{eqnarray*}
Let the set $\{i \mid x_i\in C_j\}$ be denoted by $\hat{C}_j$. It follows from the construction of the MDP $M$ and from the previously described properties that for all $i\in\{1,2,\ldots,k\}\smallsetminus \hat{C}_j$ we have that 
\begin{eqnarray*}
 \Prob[\Hit_{\mu}^i {>} 4j-3]
  & = &\Prob[\Hit_{\mu}^i {>} 4j-2]\\
  & = &\Prob[\Hit_{\mu}^i {>} 4j-1]\\ 
  & = & \Prob[\Hit_{\mu}^i {>} 4j]\\
  &= & 0.01^{j-1}\,.
\end{eqnarray*}
We distinguish four cases depending on the value of $\beta_j=\sum_{i\in \hat{C}_j}\mu(x_i)$. If $\beta_j=0$, we get
\begin{eqnarray*}
\prod_{i\in\hat{C}_j} \Prob[\Hit_{\mu}^i {>} 4j-3] &=&\prod_{i\in\hat{C}_j} 0.01^{j-1}\\
& = & 0.01^{(j-1)3}
\end{eqnarray*}
and
\begin{eqnarray*}
&&\prod_{i\in\hat{C}_j} \Prob[\Hit_{\mu}^i {>} 4j-2]\\ &=& \prod_{i\in\hat{C}_j} \Prob[\Hit_{\mu}^i {>} 4j-1]\\
&=&\prod_{i\in\hat{C}_j} \Prob[\Hit_{\mu}^i {>} 4j]\\
&=&\prod_{i\in\hat{C}_j} (0.01^{j-1}\cdot 0.505)\\
&=& 0.01^{(j-1)3}\cdot 0.505^3\,,
\end{eqnarray*}
implying that the sum
\begin{eqnarray*}
&& \prod_{i=1}^k \Prob[\Hit_{\mu}^i {>} 4j-3]
+  \prod_{i=1}^k \Prob[\Hit_{\mu}^i {>} 4j-2]\\
&+&  \prod_{i=1}^k \Prob[\Hit_{\mu}^i {>} 4j-1]
+ \prod_{i=1}^k \Prob[\Hit_{\mu}^i {>} 4j]\,
\end{eqnarray*} 
is equal to
\begin{eqnarray*} 
  (1 + 3\cdot 0.505^3)\cdot 0.01^{(j-1)k} &=& 1.386362875\cdot 0.01^{(j-1)k} \\& > & 1.2601255\cdot 0.01^{(j-1)k}\,.
\end{eqnarray*}
If $\beta_j=1$ ($\beta_j=2$, $\beta_j=3$, respectively) we obtain that the sum
\begin{eqnarray*}
&& \prod_{i=1}^k \Prob[\Hit_{\mu}^i {>} 4j-3]
+  \prod_{i=1}^k \Prob[\Hit_{\mu}^i {>} 4j-2]\\
&+&  \prod_{i=1}^k \Prob[\Hit_{\mu}^i {>} 4j-1]
+ \prod_{i=1}^k \Prob[\Hit_{\mu}^i {>} 4j]\,
\end{eqnarray*}
is equal to $1.2601255\cdot 0.01^{(j-1)k}$ ($1.505101 \cdot 0.01^{(j-1)k}$, $2.000002 \cdot 0.01^{(j-1)k}$, respectively). We thus get even the strict inequality for the cases $\beta_j=0$, $\beta_j=2$, $\beta_j=3$. Since the inequality holds in all four cases (for every $j$), it follows that $\Exp[\MHit_{\mu}]\geq B-k$. Recall the current assumption $\mu(\varphi)=0$, which implies that there is at least one $j\in \{1,2,\ldots,n\}$ such that $\beta_j=\sum_{i\in \hat{C}_j}\mu(x_i)\neq 1$. Hence, the inequality is strict for at least one $j$, therefore $\Exp[\MHit_{\mu}] > B-k$.

The constructed MSSP instance has the property that the target state $\tau$ is never reached within the first $k$ steps from the beginning and that after exactly $k$ steps, the configuration is $\bar{\mu}$ for some truth assignment $\mu$. We have proved that the expected number of steps to reach the target $\tau$ from such a configuration is always greater than or equal to $B-k$, with equality occurring if and only if $\varphi$ is true under the truth assignment $\mu$.

With these observations we are finally ready to prove the claim that a coordinated strategy $\sigma$ achieving $\Exp_\sigma[\MHit] \leq B$ exists if and only if the input formula $\Phi$ is true. For the rest of this paragraph, let us assume that $\Phi\equiv Q_1x_1Q_2x_2\ldots Q_kx_k\ \varphi$ is true. Determining the truth value of a quantified Boolean formula can be seen as a 2-player game, where the two players, called Prover and Disprover, sequentially assign truth values to the propositional variables $x_1, x_2, \ldots x_k$: if the variable is existentially quantified, then the assigned value is chosen by Prover, if the variable is universally quantified, the assigned value is chosen by Disprover. If the chosen truth values make the formula $\varphi$ true, then Prover wins, otherwise Disprover wins. The quantified Boolean formula $\Phi$ is true if and only if Prover has a winning strategy in such a game. Any such Prover's strategy can be transformed into a coordinated strategy for the constructed MSSP instance using the following correspondence: for every $i\in\{1,\ldots,k\}$, the presence of agent $i$ in one of the states $s'_{i,\ell}$ corresponds to $x_i$ not being assigned to a truth value yet, the presence of agent $i$ in one of the states $s'_{i,\top,\ell}$ (or $s'_{i,\bot,\ell}$) corresponds to $x_i$ being assigned to $1$ (or $0$, respectively), for agent $i$ such that $Q_i\equiv \exists$ the choice of action $\top$ (or $\bot$) in the $i$-th step corresponds to the Prover's choice of assigning truth value $1$ to $x_i$ (or $0$ to $x_i$, respectively), similarly, for agent $i$ such that $Q_i\equiv \forall$ the outcome of action $\bot$ in the $i$-th step corresponds to the Disprover's choice for $x_i$. Let $\sigma$ be a coordinated strategy corresponding to a winning strategy for Prover. It is now an easy observation that applying such a coordinated strategy $\sigma$ ensures that after $k$ steps from the beginning the agents will always be in a configuration $\bar{\mu}$ corresponding to some truth assignment $\mu$ satisfying $\mu(\varphi)=1$ (and thus $\Exp[\MHit_{\mu}] = B-k$), implying that $\Exp_\sigma[\MHit] = k+(B-k) \leq B$.

To prove the converse, let $\sigma$ be a (memoryless deterministic) coordinated strategy for the constructed MSSP instance such that $\Exp_\sigma[\MHit] \leq B$. As implied by the previous statements, the configuration $\bar{\mu}$ reached after exactly $k$ steps from the beginning must correspond (with probability $1$) to some truth assignment $\mu$ such that $\Exp[\MHit_{\mu}] = B-k$ (and therefore $\mu(\varphi)=1$). Using the correspondence described in the previous paragraph, $\sigma$ can be used to construct a winning strategy for Prover, witnessing that the input formula $\Phi$ is true.
\section{Proof of Theorem~3}

Let $k \geq 1$ be the number of agents. Let us fix a set of memory states $\Mem$ (finite or countably infinite), and let $\Pi_{\Mem}$ be the set of all strategy profiles for $k$ agents such that every strategy in the profile uses memory $\Mem$. 
We show that there exists $\pi^* \in \Pi_{\Mem}$ such that 
\[
    \Exp_{\pi^*}[\MHit] \ = \ \inf_{\pi' \in \Pi_{\Mem}} \Exp_{\pi'}[\MHit]\,.
\]
Clearly, there exists a sequence of profiles $\pi_1,\pi_2,\ldots$ such that 
\[
    \lim_{n \to \infty} \Exp_{\pi_n}[\MHit] \ = \ \inf_{\pi' \in \Pi_{\Mem}} \Exp_{\pi'}[\MHit]\,.
\]
For all $n \geq 1$ and $i \in \{1,\ldots,k\}$, we use $\sigma_n^i = (m_n^i,\Next_n^i,\Update_n^i)$ to denote the strategy for agent~$i$ in the profile $\pi_n$. Without restrictions, we assume there exists $m^* \in \Mem$ such that $m_n^i = m^*$ for all strategies (note that the memory elements can be permuted).

Let $\Psi = \Psi_1,\Psi_2,\ldots$ be an infinite sequence containing all elements of the set
\begin{eqnarray*}
    U & = & \{1,\ldots,k\} \times S \times \Mem \times \Act\\
      & \cup & \{1,\ldots,k\} \times  S \times \Mem \times \Act \times \Mem 
\end{eqnarray*} 
Note that $U$ is either finite or countably infinite, and hence the sequence $\Psi$ exists (if $\Mem$ is finite, then some elements of $U$ occur infinitely often in $\Psi$).

For every $n \geq 0$, we define an infinite sequence of profiles $\Upsilon_n$ inductively as follows:
\begin{itemize}
    \item $\Upsilon_0$ is the sequence $\pi_1,\pi_2,\ldots$ fixed above.
    \item Let $\Upsilon_{i} = \hat{\pi}_1,\hat{\pi}_2,\ldots$ Then
    $\Upsilon_{i+1} = \hat{\pi}_{j_1},\hat{\pi}_{j_2},\ldots$ is an infinite subsequence of $\Upsilon_{i}$ where the indexes $j_1,j_2,\ldots$ are obtained in the following way.
    Consider the element $\Psi_{i+1}$ of the sequence $\Psi$. There are two possibilities.
    \begin{itemize}
        \item $\Psi_{i+1} \equiv (i,s,m,a)$. Then, $j_1,j_2,\ldots$ are selected so that the sequence $\Next^i_{j_1}(s,m)(a), \Next^i_{j_2}(s,m)(a),\ldots$ is convergent. This is possible because every infinite sequence of real numbers in the interval $[0,1]$ contains an infinite convergent subsequence. We use $\calL_{i+1}$ to denote the corresponding limit.
        \item $\Psi_{i+1} \equiv (i,s,m,a,m')$. In this case, the indexes $j_1,j_2,\ldots$ are selected so that the sequence $\Update^i_{j_1}((s,m),a)(m'), \Update^i_{j_2}((s,m),a)(m'),$ $\ldots$ is convergent. Again, we use $\calL_{i+1}$ to denote the corresponding limit.
    \end{itemize}
\end{itemize}
Observe that if $\Psi_n = \Psi_{n'}$, then $\calL_n = \calL_{n'}$. Furthermore, for every $\Upsilon_n = \pi_1^n,\pi_2^n,\ldots$, we have that 
\[
    \lim_{j \to \infty} \Exp_{\pi_j^n}[\MHit] \ = \ \inf_{\pi' \in \Pi_{\Mem}} \Exp_{\pi'}[\MHit]\,.
\]
Consider the profile $\pi^* = (\sigma_1^*,\ldots,\sigma_k^*)$ where the strategy $\sigma_i^* = (m^*,\Next_i^*,\Update_i^*)$ is defined as follows:
\begin{itemize}
    \item $\Next_i^*(s,m)(a) = \calL_n$ where $\Psi_{n} = (i,s,m,a)$ (if there are multiple eligible~$n$'s, any of them can be chosen because all of them determine the same $\calL_n$).
    \item $\Update_i^*((s,m),a)(m') = \calL_n$ where $\Psi_{n} = (i,s,m,a,m')$.
\end{itemize}
We show that the assumption
\begin{equation}
   \Exp_{\pi^*}[\MHit] \ = \ \varepsilon + \inf_{\pi' \in \Pi_{\Mem}} \Exp_{\pi'}[\MHit]
\label{eq-eps}
\end{equation}
leads to a contradiction for an arbitrarily small fixed $\varepsilon > 0$. This implies 
\[
   \Exp_{\pi^*}[\MHit] \ = \ \inf_{\pi' \in \Pi_{\Mem}} \Exp_{\pi'}[\MHit]
\]
and the proof is finished.

To derive the contradiction, suppose that~\eqref{eq-eps} hold for some $\varepsilon > 0$.  Consider the original sequence of profiles $\pi_1,\pi_2,\ldots$ satisfying 
\[
    \lim_{n \to \infty} \Exp_{\pi_n}[\MHit] \ = \ \inf_{\pi' \in \Pi_{\Mem}} \Exp_{\pi'}[\MHit]\,.
\]
Clearly, for an arbitrarily small $\delta > 0$, there exists $n_\delta \in \Nset$ such that $\Exp_{\pi_{n}}[\MHit]$ is $\delta$-close to $\inf_{\pi' \in \Pi_{\Mem}} \Exp_{\pi'}[\MHit]$ for all $n \geq n_\delta$. Furthermore, from the definition $\pi^*$, we obtain that for an arbitrarily small $\delta>0$ there exists $m_\delta \in \Nset$ such that $\Exp_{\pi_{m}}[\MHit]$ is $\delta$-close to $\Exp_{\pi^*}[\MHit]$ for infinitely many $m\geq m_0$. However, for $\delta = \varepsilon/4$, the $n_\delta$ and $m_\delta$ cannot exist simultaneously.

\section{Proof of Theorem~4}

For the rest of this proof, let us fix an integer $n\geq 1$.  We start by  describing an instance of the Autonomous MSSP problem with $k=2$ agents. Let $M = (S, \Act, P)$ be an MDP where the set of states is $S = \{s, q, r_0, r_1, \ldots, r_{2n+2}\}$, the set of actions is $\Act = \{\top, \bot\}$ and the transition function $P$ is defined as follows: 
\begin{itemize}
    \item $P(s,\top,q)=P(r_0,\top,r_0)=1$, $P(r_{2n+2},\top,r_{2n+1})=P(r_{2n+1},\top,r_{2n})=P(r_{2n},\top,r_{2n-1})=\ldots =P(r_2,\top,r_1)=P(r_1,\top,r_0)=1$, 
    \item $P(q,\top,s)=P(q,\top,r_0)=\frac{1}{2}$, 
    \item $P(s,\bot,r_{2n+2})=\frac{3}{4}$, 
    \item $P(s,\bot,r_0)=\frac{1}{4}$,
    \item the function $P$ maps all the remaining triples to $0$. 
\end{itemize}

Let $s$, $r_{2n+2}$ be the initial states and let $T_1 = T_2 = \{r_0\}$ be the set of target states.

Let $\Mem=\{1,\ldots,n+1\}$ be the set of memory states, let $\sigma = (n+1,\Next,\Update)$ be a coordinated strategy for one agent such that:
\begin{itemize}
    \item $\Next(s,j)(\top)=1$ and $\Next(s,j)(\bot)=0$ for all $j\in\Mem\smallsetminus\{1\}$, 
    \item $\Next(s,1)(\top)=0$, $\Next(s,1)(\bot)=1$,
    \item $\Next$ returns the Dirac distribution on the set $\{\top\}$ for the remaining cases,
    \item $\Update((s,1),\top)(1)=\Update((s,1),\bot)(1)=1$,
    \item $\Update((s,j),\top)(j-1)=\Update((s,j),\bot)(j-1)=1$ for all $j\in\Mem\smallsetminus\{1\}$,
    \item $\Update((x,j),\top)(j)=\Update((x,j),\bot)(j)=1$ for all $j\in\Mem$ and $x\in S\smallsetminus\{s\}$.
\end{itemize}
Let $\pi=(\sigma, \sigma)$. Note that $\pi$ is a strategy profile of $\Pi_{n+1}$.

With the application of Theorem 3, let us fix a $\Pi_n$-optimal profile $\pi'$ such that $\Exp_{\pi'}[\MHit] = \inf_{\pi' \in \Pi_n} \Exp_{\pi'}[\MHit]$. In order to finish the proof of the first part of Theorem 4, it remains to show that $\Exp_\pi[\MHit] < \Exp_{\pi'}[\MHit]$. In the constructed MSSP instance, it holds that whenever the first agent is in the state $s$, then the second agent is in state $r_{2j}$ for some $j\in\{0,1,\ldots,n+1\}$. The strategy profile $\pi$ defined above has the property that if the first agent is in state $s$, it chooses the action $\bot$ whenever the second agent is in the state $r_2$ and chooses the action $\top$ whenever the second agent is in state $r_{2j}$ for some $j\in\{2,3,\ldots,n+1\}$. It is easy to verify that all profiles with this property achieve the same expected value of $\MHit$ and that none of the profiles of $\Pi_n$ (including $\pi'$) have this property (since with at most $n$ memory states there will always be a nonzero probability of not choosing $\bot$ when the second agent is in state $r_2$ or choosing $\bot$ before the second agent has reached $r_2$). In the following paragraphs, we show that $\pi'$ achieves strictly higher expected value of $\MHit$ compared to $\pi$, using the fact that $\pi'$ does not have the described property of $\pi$.

Let us first assume that $\pi'$ admits a nonzero probability that the first agent (being in the state $s$) chooses the action $\top$ while the second agent is in the state $r_2$. The target state $r_0$ is then first reached by some agent after exactly $2$ following steps (with probability $1$), however, choosing the action $\bot$ would lead to a strictly smaller expected number ($\frac{1}{4}\cdot 1 + \frac{3}{4}\cdot 2 = 1.75$) of steps to reach the target. Let us now assume that $\pi'$ admits a positive probability that the first agent (being in the state $s$) chooses the action $\bot$ while the second agent is in the state $r_{2j}$ for some $j\in\{2,3,\ldots,n+1\}$. The target state $r_0$ is then first reached by some agent after exactly $1$ step with probability $\frac{1}{4}$ and after exactly $2j$ steps with probability $\frac{3}{4}$, resulting in the expected number of $\frac{1}{4}\cdot 1 + \frac{3}{4}\cdot 2j = \frac{3}{2} j + \frac{1}{4}$ steps. However, choosing the action $\top$ would lead to a strictly smaller expected number of steps to reach the target (at most $\frac{1}{2}\cdot 2 + \frac{1}{2}\cdot 2j = j+1$, which is strictly less than $\frac{3}{2} j + \frac{1}{4}$ for all $j\geq 2$).

Since the strategy profile $\pi'$ does not have the described property of $\pi$, it follows that performing a sequence of successive improvements of $\pi'$ in compliance with the previous paragraph (adding new memory states may be required to implement such improvements) yields a strategy profile having the property of $\pi$ and achieving strictly smaller expected value of $\MHit$ (namely $\Exp_\pi[\MHit]$). We thus get $\Exp_\pi[\MHit] < \Exp_{\pi'}[\MHit]$, which was to be proved.

It remains to show that for every $\varepsilon > 0$ there exists an \mbox{$\varepsilon$-optimal} finite-memory strategy profile. Let $\pi=(\sigma_1,\ldots,\sigma_k)$ an \emph{optimal} strategy profile which is guaranteed to exist by Theorem~3. If $\pi$ is finite-memory, we are done immediately. Otherwise, let $\hat{\pi} = (\hat{\sigma}_1,\ldots,\hat{\sigma}_k)$ be a memoryless profile where
every $\hat{\sigma}_i$ is an optimal strategy for agent~$i$ computable by the LP of Fig.~2 in the main body of the paper. For every $\ell \in \Nset$, consider a profile $\pi[\ell] = (\sigma[\ell]_1,\ldots,\sigma[\ell]_k)$ such that every $\sigma[\ell]_i$ behaves exactly like $\sigma_i$ for the first $\ell$ steps, and then it ``switches'' to $\hat{\sigma}_i$ forever. Since $\sigma_i$ uses only finitely many memory states during the first $\ell$ steps, the strategy  $\sigma[\ell]_i$ requires only \emph{finitely many} memory states (some new memory states may be needed to implement a bounded counter that counts the steps until reaching $\ell$). It is easy to see that 
$\lim_{\ell \to \infty} \Exp_{\pi[\ell]}[\MHit] = \Exp_{\pi}[\MHit]$. Hence, for every $\varepsilon > 0$, there exists a sufficiently large $\ell$ such that $\Exp_{\pi[\ell]}[\MHit] - \Exp_{\pi}[\MHit] \leq \varepsilon$. 
\section{Randomized Profiles}

For single-agent SSP, there always exists an optimal memoryless deterministic strategy (see Fig.~2 in the main body of the paper). For the Autonomous MSSP, the role of randomization is essential. More precisely, randomized profiles may achieve strictly better performance than deterministic profiles with the same amount of memory. This is demonstrated by the following example:
\begin{center}
\begin{tikzpicture}[x=9mm, y=15mm, >=stealth', scale=1,font=\small]
        \node[state] (S)  at (0,0)  {$s_1$};
        \node[state] (S1) at (2,0 ) {};
        \node[state] (S2) at (4,0) {};
        \node[state] (S3) at (6,0)  {};
        \node[state] (S4) at (8,0)  {$\tau$};
        \node[state] (S6) at (6,-2) {$s_2$};
        \node[state] (S7) at (8,-2) {};
        \node[act]    at (1,0)  {};
        \node[act]    at (3,0)  {};
        \node[act]    at (5,0)  {};
        \node[act]    at (7,0)  {};
        \node[act]    (D) at (7,-2)  {};
        \node[act]    (A) at (8,-1)  {};
        \node[act]    (B) at (6,-1)  {};
        \node[act]    (C) at (8,.5)  {};
        \draw[tr]    (S)  --  (S1);
        \draw[tr]    (S1) --  (S2);
        \draw[tr]    (S2) --  (S3);
        \draw[tr]    (S3) --  (S4);
        \draw[tr]    (S6) --  (S7);
        \draw[tr,-]  (S6) --   (B);
        \draw[tr]    (S7) --  node[right, near end] {$\frac{9}{26}$} (S4);
        \draw[tr]    (A) -- node[above,near end]{$\frac{17}{26}$}  +(-1,0) -- (S6);
        \draw[tr]    (B) -- +(-1,0) -- node[left=1ex]{$\frac{7}{8}$}  (S6);
        \draw[tr]    (B) -- node[left,near end]{$\frac{1}{8}$} +(0,.5) -- (S4);
        \draw[tr,-]  (S4) --  (C);
        \draw[tr]    (C) -- +(.5,0) |- (S4);
        \node[right of = B, node distance=1.5ex, color=red] {$a$};
        \node[below of = D, node distance=1.5ex, color=blue] {$b$};
\end{tikzpicture}
\end{center}

\noindent
There are two agents with the initial states $s_1$ and $s_2$, and the common target state $\tau$. The first agent (starting in $s_1$) has only one strategy, leading to $\tau$ in $4$ steps with probability one. In the best memoryless deterministic profile $\pi$, agent~2 selects the action {\color{blue}$b$}. This yields $\Exp_{\pi}[\MHit] \doteq3.30769231$. However, there is a randomized memoryless profile $\hat{\pi}$ where agent~2 selects the action {\color{red}$a$} with probability $0.2626$ and action {\color{blue}$b$} with the remaining probability $0.7374$. This yields a strictly better outcome $\Exp_{\hat{\pi}}[\MHit] \doteq 3.29972213$.
\section{Proof of Theorem~5}

Let $M = (S,\Act,P)$ be an MDP, $k \geq 1$, and $\Mem$ a set of memory states of size $n \geq 1$. We define an MDP $\bar{M} = (S{\times} \Mem, \Act {\times} \Mem, \bar{P})$ such that, for all $s \in S$, $a \in \Act$, and $m,m' \in \Mem$, we have that $P(s,a,t) = p$ in $M$ iff $P((s,m),(a,m'),(t,m')) = p$ in $\bar{M}$.

Let $\sigma$ be a strategy for one agent in $M$ with memory $\Mem$. We show that there is a memoryless strategy $\bar{\sigma}$ for one agent in $\bar{M}$ such that the Markov chains $M_\sigma$ and $\bar{M}_{\bar{\sigma}}$ are identical. The strategy $\bar{\sigma} : (S{\times} \Mem) \to \Dist(\Act{\times} \Mem)$ is defined as follows:
\[
    \bar{\sigma}(s,m)(a,m') = \Next(s,m)(a) \cdot \Update((s,m),a)(m')\,.
\]

The Markov chains $M_\sigma$ and $\bar{M}_{\bar{\sigma}}$ have the same set of states $S{\times} \Mem$. Let $(s,m),(t,m') \in S{\times} \Mem$. The stochastic matrix of $M_\sigma$ assigns to the pair of states $((s,m),(t,m'))$ the probability
\begin{equation*}
   \sum_{a \in \En(s)} \Next(s,m)(a) \cdot P(s,a,t) \cdot \Update((s,m),a)(m')\,.
\end{equation*}
The stochastic matrix of  $\bar{M}_{\bar{\sigma}}$ assigns to the pair of states $((s,m),(t,m'))$ the probability
\begin{equation*}
    \sum_{(a,m') \in \En(s,m)} \bar{\sigma}(s,m)(a,m') \cdot P((s,m),(a,m'),(t,m'))
\end{equation*}
By definition of $\bar{\sigma}$, we immediately obtain that the above sums are equal. This proves the first part of Theorem~5.

The second part of Theorem~5 is proven similarly. Let  $\bar{\sigma}$ be a memoryless strategy for $\bar{M}$. We define a strategy $\sigma$ for $M$ with memory $\Mem$ as follows:
\begin{itemize}
    \item $\Next(s,m)(a) = \sum_{m' \in \Mem}\bar{\sigma}(s,m)(a,m')$
    \item $\Update((s,m),a)(m') = 
       \frac{\bar{\sigma}(s,m)(a,m')}{\sum_{m'' \in\Mem} \bar{\sigma}(s,m)(a,m'')}$
\end{itemize}
Now, it is easy to verify that the stochastic matrices of $M_\sigma$ and $\bar{M}_{\bar{\sigma}}$ assign to every pair of states the same probability. This proves the second part of Theorem~5.

\section{Proof of Theorem~6}

We prove this NP-hardness result by polynomial-time reduction from a variant of the positive 1-in-3-SAT problem where every clause contains exactly three literals and every variable occurs in exactly three clauses. An input instance of this problem is a negation-free propositional formula in conjunctive normal form (with exactly three distinct variables appearing in each clause and every variable occurring in exactly three clauses) and the task is to decide whether there exists a truth assignment such that exactly one literal is assigned the value $1$ in each clause. If such an assignment exists, we say that the formula is \emph{1-in-3-satisfiable} and we call the corresponding truth assignment \emph{1-in-3-satisfying}. To see that this restricted variant of the positive 1-in-3-SAT problem is still $\NP$-hard, notice that the $\NP$-complete RXC3 problem\footnote{See \emph{Teofilo F. Gonzales: Clustering to minimize the maximum intercluster distance, Theoretical Computer Science, Volume 38, 1985, Pages 293--306.}} is essentially the same problem, where the elements play the role of clauses, the 3-element subsets play the role of literals (propositional variables) and the presence of a 3-element subset in the subcollection (i.e., in the candidate for an exact cover) plays the role of assigning value $1$ to the corresponding propositional variable.

Let $\Phi\equiv C_0\land C_1\land\ldots\land C_{n-1}$ be an instance of the restricted variant of the positive 1-in-3-SAT problem, where each $C_j$ is a clause. In this proof, the clauses are understood as sets of three propositional variables. Without restrictions, we assume that $\bigcup_{j=0}^{n-1} C_j = \{x_0, x_1, \ldots, x_{n-1}\}\subseteq \Var$, where each $x_i\in \Var$ denotes a fresh propositional variable (the number of variables occurring in $\Phi$ must be equal to \mbox{$n\geq 3$}). In the following paragraphs, we describe a polynomial-time construction of an MSSP instance with two agents and a rational bound $B$ such that a (memoryless) strategy profile $\pi$ achieving $\Exp_\pi[\MHit] \leq B$ exists if and only if $\Phi$ is 1-in-3-satisfiable.

A key ingredient used in this construction is a \emph{gadget}, which is an MDP with two entry states $g_0$, $g_1$ and with the property that 
\begin{itemize}
  \item whenever an agent visits a state $g_0$, then it reaches the target state $\tau$ in exactly $1$ step with probability $0.5$ and in exactly $4$ steps with probability $0.5$,
  \item whenever an agent visits a state $g_1$, then it reaches the target state $\tau$ in exactly $2$ steps with probability $0.9$ and in exactly $7$ steps with probability $0.1$.
\end{itemize}
For a single agent, the expected number of steps to reach $\tau$ is equal to $2.5$, regardless of whether the current state is $g_0$ or $g_1$. However, 
\begin{itemize}
  \item when two agents simultaneously visit $g_0$, then the achieved expected number of steps to reach $\tau$ (by some agent) is equal to $1.75$,
  \item when the two agents simultaneously visit $g_1$, then the achieved expected number of steps to reach $\tau$ is equal to $2.05$,
  \item when one agent visits $g_0$ and the other visits $g_1$, then the expected number of steps to reach $\tau$ is equal to $1.6$.
\end{itemize}
We construct an MSSP instance for two agents so that there always exists an optimal strategy profile which is memoryless and deterministic. The two memoryless deterministic strategies will then correspond to two truth assignments. Using the gadget described in the previous paragraph, we enforce that an optimal memoryless deterministic strategy profile necessarily consists of two strategies that both correspond to the same truth assignment. Such pairs of strategies are called \emph{consistent}. To complete our proof, we use the same gadget to ensure the following:
\begin{itemize}
  \item If a 1-in-3-satisfying truth assignment (corresponding to the input formula $\Phi$) exists, then the pair of two consistent memoryless deterministic strategies corresponding to such an assignment forms a strategy profile $\pi$ achieving $\Exp_\pi[\MHit] = B$.
  \item If a 1-in-3-satisfying truth assignment does not exist, then every pair of consistent memoryless deterministic strategies achieves a strictly higher expected value of $\MHit$.
\end{itemize}
Let $M = (S, \Act, P)$ be an MDP where
\begin{eqnarray*}
S & = & \{\tau, s', s'', g_0, g_1, g'_1, g'_2, g'_3, g'_4, g'_5, g'_6\}\\
& \cup & \{s'_i \mid 0 \leq i \leq n-1\}\\
& \cup & \{s''_i \mid 0 \leq i \leq n-1\}\\
& \cup & \{s'_{i,0,\ell} \mid 0 \leq i \leq n-1, 0 \leq \ell \leq 8n-1\}\\
& \cup & \{s'_{i,1,\ell} \mid 0 \leq i \leq n-1, 0 \leq \ell \leq 8n-1\}\\
& \cup & \{s''_{i,0,\ell} \mid 0 \leq i \leq n-1, 0 \leq \ell \leq 8n-1\}\\
& \cup & \{s''_{i,1,\ell} \mid 0 \leq i \leq n-1, 0 \leq \ell \leq 8n-1\}\\
& \cup & \{s_{i,0} \mid 0 \leq i \leq n-1\}\\
& \cup & \{s_{i,1} \mid 0 \leq i \leq n-1\}\\
& \cup & \{s_{j,0,\ell} \mid 0 \leq j \leq n-1, 0 \leq \ell \leq 8j\}\\
& \cup & \{s_{j,1,\ell} \mid 0 \leq j \leq n-1, 0 \leq \ell \leq 8j\},\\[1ex]
\Act  & = &\{a, a_0, a_1\}
\end{eqnarray*}
and the transition function $P$ is defined as follows:
\begin{itemize}
    \item $P(\tau,a,\tau) = P(g'_1,a,\tau) = 1$,
    \item $P(g'_{\ell},a,g'_{\ell-1}) = 1$ for all $2 \leq \ell \leq 6$,
    \item $P(g_0,a,\tau) = P(g_0,a,g'_3) = 0.5$,
    \item $P(g_1,a,g'_1) = 0.9$,
    \item $P(g_1,a,g'_6) = 0.1$,
    \item $P(s',a,s'_i) = \frac{1}{n}$ for all $0 \leq i \leq n-1$,
    \item $P(s'_i,a_0,s'_{i,0,0}) = P(s'_i,a_1,s'_{i,1,0}) = 1$ for all $0 \leq i \leq n{-}1$,
    \item $P(s'_{i,0,8i},a,g_0) = P(s'_{i,1,8i},a,g_1) = 1-\frac{1}{8n^2}$ for all $0 \leq i \leq n-1$,
    \item $P(s'_{i,0,8i},a,s'_{i,0,8i+1}) = P(s'_{i,1,8i},a,s'_{i,1,8i+1}) = \frac{1}{8n^2}$ for all $0 \leq i \leq n-1$,
    \item $P(s'_{i,0,\ell},a,s'_{i,0,\ell+1}) = P(s'_{i,1,\ell},a,s'_{i,1,\ell+1}) = 1$ for all $0 \leq i \leq n-1$ and $0 \leq \ell \leq 8n-2$ such that $\ell \neq 8i$,
    \item $P(s'_{i,0,8n-1},a,s_{i,0}) = P(s'_{i,1,8n-1},a,s_{i,1}) = 1$ for all $0 \leq i \leq n-1$,
    \item $P(s'',a,s''_i) = \frac{1}{n}$ for all $0 \leq i \leq n-1$,
    \item $P(s''_i,a_0,s''_{i,0,0}) = P(s''_i,a_1,s''_{i,1,0}) = 1$ for all $0 \leq i \leq n-1$,
    \item $P(s''_{i,0,8i},a,g_1) = P(s''_{i,1,8i},a,g_0) = 1-\frac{1}{8n^2}$ for all $0 \leq i \leq n-1$,
    \item $P(s''_{i,0,8i},a,s''_{i,0,8i+1}) = P(s''_{i,1,8i},a,s''_{i,1,8i+1}) = \frac{1}{8n^2}$ for all $0 \leq i \leq n-1$,
    \item $P(s''_{i,0,\ell},a,s''_{i,0,\ell+1}) = P(s''_{i,1,\ell},a,s''_{i,1,\ell+1}) = 1$ for all $0 \leq i \leq n-1$ and $0 \leq \ell \leq 8n-2$ such that $\ell \neq 8i$,
    \item $P(s''_{i,0,8n-1},a,s_{i,0}) = P(s''_{i,1,8n-1},a,s_{i,1}) = 1$ for all $0 \leq i \leq n-1$,
    \item $P(s_{i,0},a,s_{j,0,0}) = P(s_{i,1},a,s_{j,1,0}) = \frac{1}{3}$ for all $0 \leq i \leq n-1$ and $0 \leq j \leq n-1$ such that $x_i\in C_j$,
    \item $P(s_{j,0,\ell},a,s_{j,0,\ell+1}) = P(s_{j,1,\ell},a,s_{j,1,\ell+1}) = 1$ for all $0 \leq j \leq n-1$ and $0 \leq \ell \leq 8j-1$,
    \item $P(s_{j,0,8j},a,g_0) = P(s_{j,1,8j},a,g_1) = 1$ for all $0 \leq j \leq n-1$,
    \item the function $P$ maps all the remaining triples to $0$.
\end{itemize}

Let $T_1 = T_2 = \{\tau\}$ be the (shared) set of target states, and let $\iota_1 = s'$, $\iota_2 = s''$ be the initial states of agent~$1$ and agent~$2$. Furthermore, let 
\[
  B = \frac{10240n^6 + 5760n^5 + 2944n^4 - 96n^2 + 60n + 59}{3840n^5}\,.
\]

In this instance of the MSSP problem for two agents, there is always only the action $a$ enabled in every state of the MDP~$M$ except for the states $s'_i$, $s''_i$ which are visited after exactly one step from the beginning. This implies that, for this particular MSSP instance, memoryless profiles are sufficient to achieve all values of $\Exp_\pi[\MHit]$ that are achievable by general strategy profiles, because storing information about the history obviously gives the agents no advantage with respect to the achieved value of $\Exp_\pi[\MHit]$. Moreover, there is no advantage in using randomized strategy profiles in this instance, as the value of $\Exp_\pi[\MHit]$ achieved by a randomized strategy profile is always equal to some weighted arithmetic mean of the values of $\Exp_\pi[\MHit]$ achievable by deterministic strategy profiles (i.e., it is possible to successfully derandomize the strategies). Therefore, there always exists an optimal strategy profile 
which is memoryless and deterministic.

The MDP $M$ is designed so that each agent is forced to take action~$a$ in its initial state, changing the agent's state randomly (with uniform distribution) to some of the states $s'_i$ ($s''_i$, respectively), where the agent has to choose between the two available actions $a_0$ and $a_1$. The choice of $a_0$ ($a_1$, respectively) corresponds to the assignment of Boolean value $0$ ($1$, respectively) to the propositional variable $x_i$. If agent~$1$ chooses $a_0$ ($a_1$, respectively), then with probability $1-\frac{1}{8n^2}$ this agent reaches the gadget's entry state $g_0$ ($g_1$, respectively) after exactly $8i+3$ steps from the beginning. With the remaining probability $\frac{1}{8n^2}$, this agent reaches the state $s_{i,0}$ ($s_{i,1}$, respectively) after exactly $8n+2$ steps from the beginning. If agent $2$ chooses $a_0$ ($a_1$, respectively), then with probability $1-\frac{1}{8n^2}$ this agent reaches the gadget's entry state $g_1$ ($g_0$, respectively) after exactly $8i+3$ steps from the beginning, and with the remaining probability $\frac{1}{8n^2}$ this agent reaches the state $s_{i,0}$ ($s_{i,1}$, respectively) after exactly $8n+2$ steps from the beginning. Notice the swapped gadget's entry states for agent $2$; the purpose of this swap is to ensure the consistency of the optimal deterministic strategies (where consistency means that agent~$1$ chooses $a_1$ in $s'_i$ if and only if agent~$2$ chooses $a_1$ in~$s''_i$).

The construction has the property that each of the two agents can reach the target state $\tau$ only through the gadget. Whenever one of the agents (let us denote this agent $\alpha$) reaches one of the states $g_0$, $g_1$, then the other agent~$\beta$ either reaches one of the states $g_0$, $g_1$ in the same step as agent~$\alpha$, or it does not reach $g_0$, $g_1$ before agent~$\alpha$ reaches the target state~$\tau$. Furthermore, observe that if some agent does not reach the target~$\tau$ within the first $8n+2$ steps (this happens with probability $\frac{1}{8n^2}$, when the agent first reaches one of the states $s_{i,0}$ or $s_{i,1}$), then it enters the gadget's state $g_0$ or $g_1$ from some of the states $s_{j,0,8j}$ or $s_{j,1,8j}$ (depending on whether the action taken in the second step is $a_0$ or $a_1$): for every $j\in\{0,\ldots,n-1\}$, the probability that the agent reaches one of the two states $s_{j,0,8j}$, $s_{j,1,8j}$ is equal to $\frac{1}{8n^2}\cdot \frac{1}{n}$. If the agent reaches one of the two states $s_{j,0,8j}$, $s_{j,1,8j}$ for some $j\in\{0,\ldots,n-1\}$, then the conditional probability that the agent has previously reached one of the two states $s_{i,0}$, $s_{i,1}$ is equal to $\frac{1}{3}$ for all $i$ such that $x_i\in C_j$ (and is equal to $0$ for the rest).

Let $\pi=(\sigma_1, \sigma_2)$ be a memoryless deterministic strategy profile. Let $\zeta_1$ be the expected value of $\MHit$ under the condition that agent $1$ (or $2$, respectively) reaches the target $\tau$ in at most $8n+2$ steps and agent $2$ ($1$, respectively) reaches the target $\tau$ in more than $8n+2$ steps. In this case, the value of $\MHit$ is equal to the number of steps taken by agent $1$ ($2$, respectively) and the (conditional) expected value of $\MHit$ may thus be expressed easily as 
\[
  \zeta_1 = \sum_{i=0}^{n-1}\frac{1}{n}(8i+3+2.5)
\]  
Notice that the value of $\zeta_1$ is independent of $\pi$. Let $\zeta_2(\pi)$ be the expected value of $\MHit$ under the condition that both agents reach the target $\tau$ in at most $8n+2$ steps. In this case, we further distinguish the cases when one agent reaches the gadget strictly earlier than the other one, and when both agents enter the gadget simultaneously, obtaining 
\begin{eqnarray*}
  \zeta_2(\pi) & = & \zeta_2' + \zeta_2''(\pi) \\
  &=& \sum_{i=0}^{n-1}2\frac{1}{n}\frac{n-i-1}{n}(8i+3+2.5) \\
  && + \sum_{i=0}^{n-1}\left(\frac{1}{n}\right)^2(8i+3+\chi_i(\pi))
\end{eqnarray*}  
where $\chi_i(\pi)$ expresses the conditional expected number of steps taken to reach the target $\tau$ by some agent from the point when both agents first enter the gadget (simultaneously) after exactly $8i+3$ steps from the beginning. 
We have that
\[
   \chi_i(\pi) = 
\begin{cases}
   1.6 & \mbox{if } \sigma_1(s'_i)(a_1)=\sigma_2(s''_i)(a_1),\\
   1.75 & \mbox{if } \sigma_1(s'_i)(a_1)=0 \mbox{ and } \sigma_2(s''_i)(a_1)=1, \\
   2.05 & \mbox{if } \sigma_1(s'_i)(a_1)=1 \mbox{ and } \sigma_2(s''_i)(a_1)=0\,.
\end{cases}
\]
Notice that the value of $\zeta_2'$ is independent of $\pi$. Similarly, let $\zeta_3(\pi)$ be the expected value of $\MHit$ under the condition that both agents reach the target $\tau$ in more than $8n+2$ steps. Again, we further distinguish the cases when one agent reaches the gadget strictly earlier than the other one and when both agents enter the gadget simultaneously, obtaining 
\begin{eqnarray*}
 \zeta_3(\pi) & = &  \zeta_3' + \zeta_3''(\pi)\\
 & = & \sum_{j=0}^{n-1}2\frac{1}{n}\frac{n-j-1}{n}(8n+4+8j+2.5)\\
 &&  + \sum_{j=0}^{n-1}\left(\frac{1}{n}\right)^2(8n+4+8j+\chi'_j(\pi))
\end{eqnarray*} 
where $\chi'_j(\pi)$ expresses the conditional expected number of steps taken to reach the target $\tau$ by some agent from the point when both agents first enter the gadget (simultaneously) after exactly $8n+4+8j$ steps from the beginning. Let $j\in\{0,\ldots,n-1\}$ be an index of a clause $C_j = \{x_{i_1}, x_{i_2}, x_{i_3}\}$ (where $i_1, i_2, i_3 \in \{0,\ldots,n-1\}$), and let $k'_j(\pi)$ ($k''_j(\pi)$, respectively) be the number of indices $i\in\{i_1, i_2, i_3\}$ satisfying $\sigma_1(s'_i)(a_1)=1$ ($\sigma_2(s''_i)(a_1)=1$, respectively). The value of $\chi'_j(\pi)$ may be expressed as 
\begin{eqnarray*}
  \chi'_j(\pi) & = & \frac{3-k'_j(\pi)}{3}\frac{3-k''_j(\pi)}{3}\cdot 1.75\\
   &&  +\ \frac{k'_j(\pi)}{3}\frac{k''_j(\pi)}{3}\cdot 2.05\\ 
   &&  + \left(\frac{k'_j(\pi)}{3}\frac{3-k''_j(\pi)}{3}+\frac{3-k'_j(\pi)}{3}\frac{k''_j(\pi)}{3}\right)\cdot 1.6
\end{eqnarray*}
It always holds that $1.6\leq\chi'_j(\pi)\leq 2.05$. Using the law of total expectation, we can now express the expected value of $\MHit$ as 
\begin{eqnarray*}
  \Exp_\pi[\MHit] & = & 2\left(1-\frac{1}{8n^2}\right)\frac{1}{8n^2}\zeta_1
   + \left(1-\frac{1}{8n^2}\right)^2\zeta_2(\pi)\\
  && + \left(\frac{1}{8n^2}\right)^2\zeta_3(\pi)\\ 
  & = &  2\left(1-\frac{1}{8n^2}\right)\frac{1}{8n^2}\zeta_1
    + \left(1-\frac{1}{8n^2}\right)^2\zeta_2'\\
  && + \left(1-\frac{1}{8n^2}\right)^2\zeta_2''(\pi)
   + \left(\frac{1}{8n^2}\right)^2\zeta_3'\\
  && + \left(\frac{1}{8n^2}\right)^2\zeta_3''(\pi)
\end{eqnarray*}  

For the rest of this paragraph, assume that the memoryless deterministic strategy profile $\pi=(\sigma_1, \sigma_2)$ is \emph{not} consistent, i.e., there exists $i\in\{0,\ldots,n-1\}$ such that $\sigma_1(s'_i)(a_1)\neq\sigma_2(s''_i)(a_1)$. Let $\pi'=(\sigma_1', \sigma_2')$ be the strategy profile where both agents always choose the action $a_1$ in the second step (i.e., $\sigma_1(s'_i)(a_1)=\sigma_2(s''_i)(a_1)=1$ for all $i\in\{0,\ldots,n-1\}$). Let us compare the achieved expected values of $\MHit$. We have that $\Exp_\pi[\MHit] - \Exp_{\pi'}[\MHit]$ is equal to 
{\small
\begin{eqnarray*}
& & 
2\left(1-\frac{1}{8n^2}\right)\frac{1}{8n^2}\zeta_1 + \left(1-\frac{1}{8n^2}\right)^2\zeta_2'\\
&& + \left(1-\frac{1}{8n^2}\right)^2\zeta_2''(\pi)
 + \left(\frac{1}{8n^2}\right)^2\zeta_3'\\
&& + \left(\frac{1}{8n^2}\right)^2\zeta_3''(\pi)
 - 2\left(1-\frac{1}{8n^2}\right)\frac{1}{8n^2}\zeta_1\\
&& - \left(1-\frac{1}{8n^2}\right)^2\zeta_2'
 - \left(1-\frac{1}{8n^2}\right)^2\zeta_2''(\pi')\\
&& - \left(\frac{1}{8n^2}\right)^2\zeta_3' 
- \left(\frac{1}{8n^2}\right)^2\zeta_3''(\pi')\\
& = & 
 \left(1-\frac{1}{8n^2}\right)^2\left(\zeta_2''(\pi) - \zeta_2''(\pi')\right)
  + \left(\frac{1}{8n^2}\right)^2\left(\zeta_3''(\pi) - \zeta_3''(\pi')\right)\\
& = & \left(1-\frac{1}{8n^2}\right)^2\left(\sum_{i=0}^{n-1}\left(\frac{1}{n}\right)^2(8i+3+\chi_i(\pi))\right)\\
 & & - \left(1-\frac{1}{8n^2}\right)^2\left(\sum_{i=0}^{n-1}\left(\frac{1}{n}\right)^2(8i+3+1.6)\right)\\
&& + \left(\frac{1}{8n^2}\right)^2\left(\sum_{j=0}^{n-1}\left(\frac{1}{n}\right)^2(8n+4+8j+\chi'_j(\pi))\right)\\ 
&& - \left(\frac{1}{8n^2}\right)^2\left(\sum_{j=0}^{n-1}\left(\frac{1}{n}\right)^2(8n+4+8j+2.05)\right)\\
& \geq & \left(1-\frac{1}{8n^2}\right)^2 \left(\frac{1}{n}\right)^2 (1.75-1.6)\\
&& + \left(\frac{1}{8n^2}\right)^2 n\left(\frac{1}{n}\right)^2 (1.6-2.05)\\
& \geq & \left(1-\frac{1}{8}\right)^2 \frac{1}{n^2}\cdot 0.15 - \frac{1}{64n^5} \cdot 0.45\\
& \geq & \frac{1}{n^2}\left(\frac{49}{64} \cdot 0.15 - \frac{1}{64n^3} \cdot 0.45\right)\\
& \geq & \frac{1}{n^2}\left(\frac{49}{64} \cdot 0.15 - \frac{1}{64} \cdot 0.45\right) > 0
\end{eqnarray*}}%

We thus get that $\Exp_\pi[\MHit] > \Exp_{\pi'}[\MHit]$, which implies that $\pi$ is not optimal. It follows that every optimal strategy profile for the constructed instance of the MSSP problem must be consistent.

Recall that we aim to prove that a strategy profile $\pi$ achieving $\Exp_\pi[\MHit] \leq B$ exists if and only if $\Phi$ is \mbox{1-in-3-satisfiable}. For the rest of this paragraph, assume that $\Phi$ is 1-in-3-satisfiable. Let $\nu: \Var \to \{0,1\}$ be a corresponding 1-in-3-satisfying truth assignment and let $\pi=(\sigma_1, \sigma_2)$ be the memoryless deterministic strategy profile such that $\sigma_1(s'_i)(a_1)=\sigma_2(s''_i)(a_1)=\nu(x_i)$ (implying that $\sigma_1(s'_i)(a_0)=\sigma_2(s''_i)(a_0)=1-\nu(x_i)$). The achieved expected value of $\MHit$ may be expressed as 
{\small
\begin{eqnarray*}
 \Exp_\pi[\MHit] & = & 2\left(1-\frac{1}{8n^2}\right)\frac{1}{8n^2}\zeta_1\\
 && + \left(1-\frac{1}{8n^2}\right)^2\zeta_2' + \left(1-\frac{1}{8n^2}\right)^2\zeta_2''(\pi)\\
 && + \left(\frac{1}{8n^2}\right)^2\zeta_3' + \left(\frac{1}{8n^2}\right)^2\zeta_3''(\pi)\\
 & = & 2\left(1-\frac{1}{8n^2}\right)\frac{1}{8n^2}\sum_{i=0}^{n-1}\frac{1}{n}(8i+3+2.5) \\
 && + \left(1-\frac{1}{8n^2}\right)^2 \sum_{i=0}^{n-1}2\frac{1}{n}\frac{n-i-1}{n}(8i+3+2.5)\\ 
 && + \left(1-\frac{1}{8n^2}\right)^2 \sum_{i=0}^{n-1}\left(\frac{1}{n}\right)^2(8i+3+\chi_i(\pi))\\
 && + \left(\frac{1}{8n^2}\right)^2 \sum_{j=0}^{n-1}2\frac{1}{n}\frac{n-j-1}{n}(8n+4+8j+2.5)\\
 && + \left(\frac{1}{8n^2}\right)^2 \sum_{j=0}^{n-1}\left(\frac{1}{n}\right)^2(8n+4+8j+\chi'_j(\pi))
\end{eqnarray*}}%

\noindent
where all $\chi_i(\pi) = 1.6$ and all 
{\small
\begin{eqnarray*}
 \chi'_j(\pi) & = & \frac{3-1}{3}\frac{3-1}{3}\cdot 1.75 + \frac{1}{3}\frac{1}{3}\cdot 2.05\\
 && + \left(\frac{1}{3}\frac{3-1}{3}+\frac{3-1}{3}\frac{1}{3}\right)\cdot 1.6\\
 &  = &\frac{103}{60}
\end{eqnarray*}}%

\noindent
The simplified form 
{\small
\begin{eqnarray*}
\Exp_\pi[\MHit] &= &\frac{10240n^6 + 5760n^5 + 2944n^4 - 96n^2 + 60n + 59}{3840n^5}\\
& = & B
\end{eqnarray*}}%

\noindent
can be obtained using standard techniques.

To prove the converse, let $\pi=(\sigma_1, \sigma_2)$ be an arbitrary strategy profile achieving $\Exp_\pi[\MHit] \leq B$. Applying the aforementioned observations, we assume without loss of generality that the strategies $\sigma_1$, $\sigma_2$ are memoryless, deterministic and consistent. Let $\nu: Var \to \{0,1\}$ be the truth assignment such that $\nu(x_i)=\sigma_1(s'_i)(a_1)$ for all $i\in\{0,\ldots,n-1\}$ (and $\nu(y)=0$ for all remaining variables $y\in Var$). Let $j\in\{0,\ldots,n-1\}$ be an arbitrary index of a clause $C_j = \{x_{i_1}, x_{i_2}, x_{i_3}\}$ of the input formula $\Phi$ and let $k$ be the number of indices $i\in\{i_1, i_2, i_3\}$ satisfying $\nu(x_i) = 1$. It follows that 
\[
   \chi'_i(\pi) = 
\begin{cases}
   1.75 = \frac{105}{60} & \mbox{if } k=0,\\[1ex]
   \frac{103}{60} & \mbox{if } k=1,\\[1ex]
   \frac{109}{60} & \mbox{if } k=2,\\[1ex]
   2.05 = \frac{123}{60} & \mbox{if } k=3.
\end{cases}
\]
Using the expressions of the previous paragraphs, it is clear that this number $k$ must be equal to~$1$ (for all $j$), otherwise $\Exp_\pi[\MHit]$ would exceed the given bound $B$. Hence, $\nu$ assigns the value $1$ to exactly one variable in each clause of the input formula $\Phi$, which means that $\Phi$ is \mbox{1-in-3-satisfiable}.

\section{Proof of Theorem~7}

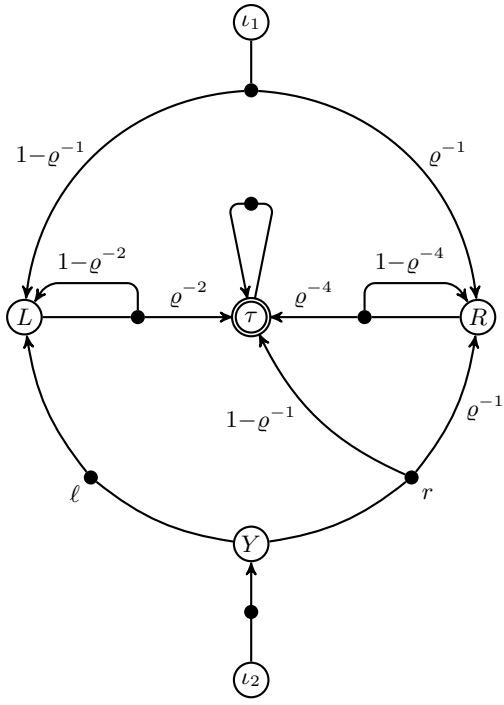
\begin{figure}
\begin{center}
\begin{tikzpicture}[x=3cm, y=3cm, >=stealth', scale=1,font=\small]
    \node[state,double] (T)  at (0,0)     {$\tau$};
    \node[state]        (R)  at (0:1)     {$R$};
    \node[state]        (L)  at (180:1)   {$L$};
    \node[state]        (Y)  at (270:1)   {$Y$};
    \node[state]        (X)  at (270:1.6) {$\iota_2$};
    \node[state]        (A)  at (90:1.3)  {$\iota_1$};
    \node[act]          (x)  at (270:1.3) {};
    \node[act]          (a)  at (90:1)   {};
    \node[act]          (y1) at (225:1)  {};
    \node at (225:1.1) {$\ell$};
    \node[act]          (y2) at (315:1)   {};
    \node at (315:1.1)   {$r$};
    \node[act]          (r)  at (0:.5)   {};
    \node[act]          (l)  at (180:.5)   {};
    \node[act]          (t)  at (90:.5)   {};
    \draw[tr]              (X)  --  (Y);
    \draw[tr,-]            (A)  --  (a);
    \draw[tr,-]            (R)  --  (r);
    \draw[tr]              (r)  --  node[above] {$\varrho^{-4}$} (T);
    \draw[tr]              (r)  -- +(-0,.15) -- node[above] {$1{-}\varrho^{-4}$}
                                +(.4,.15) -- (R);
    \draw[tr,-]            (L)  --  (l);
    \draw[tr]              (l)  -- node[above] {$\varrho^{-2}$}  (T);
    \draw[tr]              (l)  -- +(-0,.15) -- node[above] {$1{-}\varrho^{-2}$}
                                +(-.4,.15) -- (L);
    \draw[tr]    (Y)  edge [bend left=15,-] (y1) 
                 (y1) edge [bend left=15] (L); 
    \draw[tr]    (Y)  edge [bend right=15,-] (y2) 
                 (y2) edge [bend right=15] node[right] {$\varrho^{-1}$} (R); 
    \draw[tr]    (a)  edge [bend right=40] node[left] {$1{-}\varrho^{-1}$} (L); 
    \draw[tr]    (a)  edge [bend left=40] node[right=1ex] {$\varrho^{-1}$}(R);
    \draw[tr]    (y2) edge [bend left=20] node[left=1ex] {$1{-}\varrho^{-1}$} (T);
    \draw[tr]    (T) -- +(.1,.5) -- +(-.1,0.5) -- (T);     
\end{tikzpicture}
\end{center}
\caption{The price of autonomy can be arbitrarily large even for two agents.}
\label{fig:price}
\end{figure}

Consider the MDP of Fig.~\ref{fig:price} parameterized by $\varrho >1$, where $\iota_1,\iota_2$ are the initial states of agent~$1$ and agent~$2$ and $T_1 = T_2 = \{\tau\}$. Note that agent~$1$ can reach only the states $\iota_1,L,R,\tau$ with only one enabled action. Agent~2 can make decisions only in the state~$Y$. 

Consider a memoryless coordinated strategy $\sigma$ such that agent~$2$ in the state $Y$ performs either the action $\ell$ or $r$ depending on whether agent~$1$ is in the state $R$ or $L$, respectively. By evaluating $\Exp_\sigma[\MHit]$, we obtain that $\Exp_\sigma[\MHit] \leq c \cdot \varrho$ for a suitable constant $c$, i.e., $\Exp_\sigma[\MHit]$ is \emph{asymptotically linear} in~$\varrho$.

Now consider an autonomous strategy profile $\pi$ for two agents. Agent~$2$ may choose between the actions $\ell$ or $r$ at the state $Y$, but now \emph{independently} of the current position of agent~$1$. Note that as $\varrho$ increases, agent~$1$ enters the state~$L$ with probability close to~$1$. If agent~$2$ select the action~$\ell$, then $\Exp_\pi[\MHit] \geq d \cdot \varrho^2$ for a suitable constant $d$. A carefull analysis reveals that the same holds if agent~$2$ selects the action~$r$. Hence, $\Exp_\pi[\MHit]$ is \emph{asymptotically quadratic} in~$\varrho$ for \emph{every} strategy of agent~$2$. This implies that the price of autonomy can be arbitrarily large by choosing a suitably large~$\varrho$. 
 
Let us note that the argument can be adapted so that it \emph{avoids} using arbitrarily small probabilities in the underlying MPD. A small value of $\varrho^{-1}$ can be ``emulated'' by inserting $O(\log(\varrho))$ auxiliary states and actions, and using only transition probabilities equal to $1/2$.   

\section{Additional Experimental Results}

\begin{figure}[t]
  \centering

  \includegraphics[width=8cm]{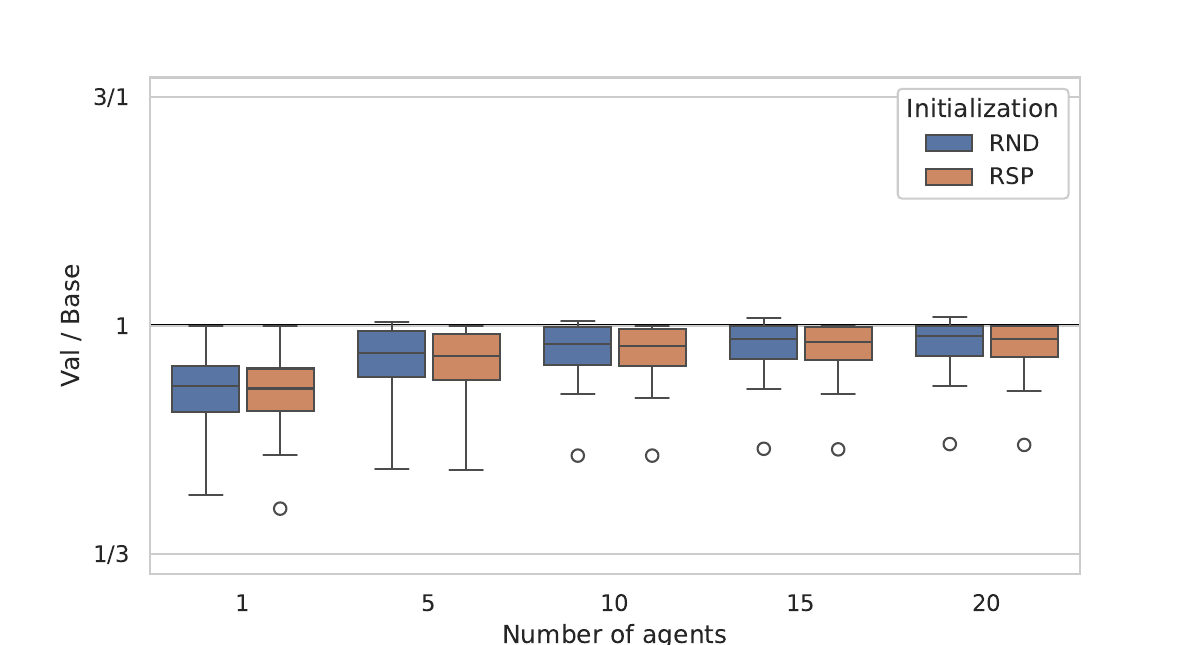}
  
  \caption{We report the $\Val_{\RND}/\Base_{\SP}$ (blue) and $\Val_{\RSP}/\Base_\SP$ (brown) ratios. The points below the line  $y=1$ correspond to benchmarks where the ratio is smaller than~$1$, i.e., the profiles computed by \mbox{\Multihit} outperform the baseline.}
  \label{fig:results_gd_sp}
\end{figure}

Figure~\ref{fig:results_gd_sp} shows the comparison of the values of strategy profiles synthesized by \mbox{\Multihit} compared to the baseline strategy profile $\pi_\SP$ that for each agent uses the graph-theoretic shortest path. We executed \mbox{\Multihit} with two initial parameter settings representing a random profile $\pi_{\RND}$ and a precomputed profile $\pi_{\RSP}$ which is a ``randomized version'' of $\pi_{\SP}$.

\begin{figure*}[t]
  \centering

  \includegraphics[width=\linewidth]{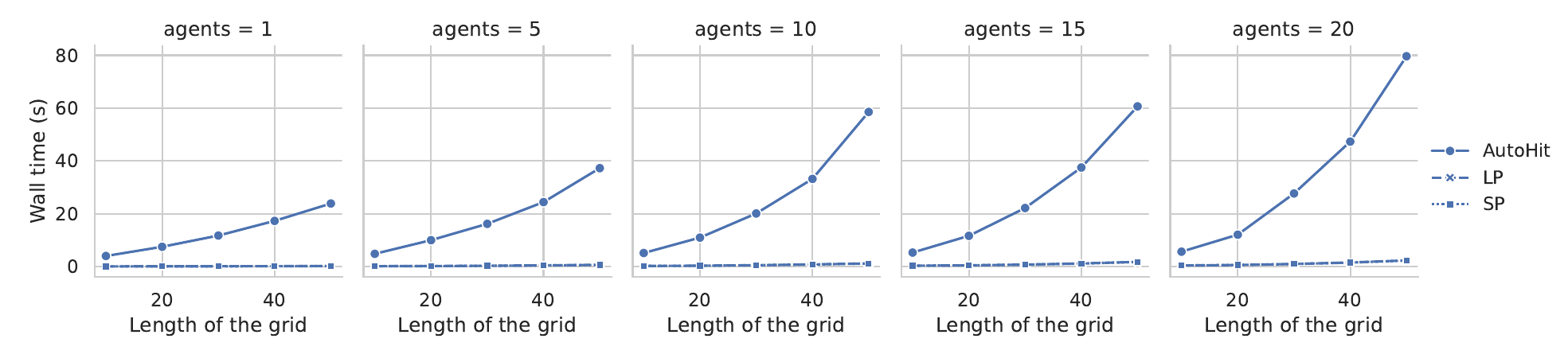}
  
  \caption{Average execution times of \mbox{\Multihit} and of the compared baselines.}
  \label{fig:results_autonomous}
\end{figure*}

Figure~\ref{fig:results_autonomous} shows the execution times of \mbox{\Multihit} and the execution times of the two baselines: \LP~and \SP. All the times are wall-clock times.

Table~\ref{tbl:autohit_table} shows all the individual values of strategy profiles for 20 agents of all compared algorithms for synthesis of autonomous strategy profiles.
\begin{table*}
\setlength\tabcolsep{8pt}
\begin{tabular}{rrrrrrr}
\toprule
Columns & Seed & $\Val_\LP$ & $\Val_\SP$ & $\Val_\RND$ & $\Val_\RLP$ & $\Val_\RSP$ \\
\midrule
10 & 1 & 11.00 & 9.07 & 9.10 & 9.07 & 9.07 \\
10 & 2 & 11.00 & 10.01 & 10.19 & 10.13 & 9.93 \\
10 & 3 & 11.00 & 9.00 & 9.01 & 9.00 & 9.00 \\
10 & 4 & 11.00 & 9.05 & 9.07 & 9.11 & 9.05 \\
10 & 5 & 11.00 & 9.12 & 9.14 & 9.12 & 9.12 \\
10 & 6 & 9.00 & 9.00 & 9.00 & 9.00 & 9.00 \\
10 & 7 & 11.00 & 10.02 & 10.00 & 10.10 & 9.93 \\
10 & 8 & 11.00 & 9.07 & 9.09 & 9.07 & 9.07 \\
10 & 9 & 13.01 & 11.35 & 11.19 & 11.24 & 11.23 \\
10 & 10 & 13.00 & 11.57 & 10.95 & 11.01 & 10.86 \\
\midrule
20 & 1 & 23.00 & 22.31 & 21.18 & 21.05 & 21.03 \\
20 & 2 & 23.00 & 19.52 & 19.80 & 20.49 & 19.52 \\
20 & 3 & 21.00 & 25.07 & 21.02 & 21.00 & 21.00 \\
20 & 4 & 25.00 & 23.86 & 22.09 & 22.79 & 21.50 \\
20 & 5 & 23.00 & 27.26 & 22.81 & 23.00 & 22.21 \\
20 & 6 & 21.00 & 19.08 & 19.16 & 19.08 & 19.08 \\
20 & 7 & 23.00 & 21.52 & 21.00 & 21.00 & 20.83 \\
20 & 8 & 21.15 & 20.76 & 21.17 & 21.13 & 20.76 \\
20 & 9 & 19.14 & 19.14 & 19.24 & 19.14 & 19.14 \\
20 & 10 & 23.00 & 21.92 & 21.18 & 21.02 & 20.98 \\
\midrule
30 & 1 & 35.00 & 34.69 & 33.02 & 32.38 & 32.46 \\
30 & 2 & 33.39 & 43.04 & 33.10 & 32.27 & 32.27 \\
30 & 3 & 37.00 & 31.21 & 32.52 & 31.71 & 31.21 \\
30 & 4 & 35.00 & 35.46 & 33.25 & 32.68 & 32.42 \\
30 & 5 & 31.00 & 33.84 & 31.12 & 31.00 & 30.98 \\
30 & 6 & 31.04 & 32.76 & 31.29 & 31.04 & 31.04 \\
30 & 7 & 37.00 & 43.91 & 33.40 & 33.23 & 33.23 \\
30 & 8 & 33.00 & 32.98 & 31.96 & 31.13 & 31.07 \\
30 & 9 & 35.00 & 61.17 & 34.63 & 35.00 & 34.49 \\
30 & 10 & 31.02 & 36.15 & 31.19 & 31.02 & 31.02 \\
\midrule
40 & 1 & 45.54 & 48.29 & 45.36 & 44.65 & 44.61 \\
40 & 2 & 45.00 & 47.72 & 43.75 & 43.14 & 43.19 \\
40 & 3 & 43.00 & 53.38 & 43.15 & 43.00 & 42.62 \\
40 & 4 & 45.02 & 47.20 & 44.88 & 43.57 & 44.11 \\
40 & 5 & 43.00 & 44.24 & 41.14 & 41.00 & 40.99 \\
40 & 6 & 43.09 & 57.98 & 43.39 & 42.40 & 42.40 \\
40 & 7 & 49.00 & 57.01 & 46.04 & 45.05 & 44.90 \\
40 & 8 & 43.00 & 40.33 & 41.83 & 40.81 & 40.33 \\
40 & 9 & 50.12 & 61.13 & 49.06 & 47.92 & 48.54 \\
40 & 10 & 41.00 & 42.17 & 41.08 & 41.00 & 40.87 \\
\midrule
50 & 1 & 57.26 & 59.45 & 57.01 & 55.74 & 55.76 \\
50 & 2 & 55.00 & 60.35 & 54.14 & 53.15 & 52.94 \\
50 & 3 & 59.05 & 66.45 & 57.51 & 56.86 & 57.89 \\
50 & 4 & 58.81 & 74.90 & 60.30 & 58.81 & 59.17 \\
50 & 5 & 57.00 & 61.27 & 55.35 & 55.08 & 55.19 \\
50 & 6 & 57.00 & 68.82 & 55.59 & 54.77 & 53.87 \\
50 & 7 & 57.00 & 67.35 & 56.83 & 55.65 & 55.95 \\
50 & 8 & 55.00 & 51.33 & 53.18 & 52.34 & 51.31 \\
50 & 9 & 64.81 & 68.01 & 66.29 & 64.06 & 63.66 \\
50 & 10 & 53.06 & 54.58 & 53.33 & 52.88 & 52.57 \\
\bottomrule
\end{tabular}

\caption{Values of strategy profiles for 20 agents of all compared algorithms for synthesis of autonomous strategy profiles.}
\label{tbl:autohit_table}
\end{table*}
\section{The Influence of the Choice of $\gamma$}

In this section, we discuss the influence of the choice of the hyperparameter $\gamma$.

\paragraph{Benchmarks.}
For evaluation, we use the same family of grid-like city benchmarks as in the main experiments. The only difference is that we use only benchmarks of size $l = 30$.
Hence, we use 10 benchmarks instead of the 50 in the main experiments.

\paragraph{Setup.}
For this experiment, we used a Linux machine with an 11th~Gen Intel Core i7--1165G7 2.8~GHz CPU, 15~GiB of RAM.
GPUs were not used in our experiments.
For each benchmark, we ran \mbox{\Multihit} $5$ times. 
We ran the algorithm \mbox{\Multihit} for each number of agents
$k \in \{ 5, 10, 15, 20 \}$.
All hyperparameters except $\gamma$ are set to the same values as in the main experiments.

For a fixed MDP $M$ and $r \geq 0$, let $\gamma_r$ denote the $r$ multiple of the number of states in $M$.
Recall that in the main experiments $\gamma$ was set to $\gamma_1$. 
Here, we use values $\gamma_r$ for $r \in \{0.2, 0.5, 0.8, 1.0, 2.0\}$.
Note that since the number of states in our benchmarks is always divisible by $5$, all values $\gamma_r$ considered here are integers.

As in the main text, we use two initial parameter settings.
One representing a random profile $\pi_{\RND}$, and the other a precomputed profile $\pi_{\RLP}$ which is a ``randomized version'' of $\pi_{\LP}$.

\begin{table*}[tb]
\centering
\begin{tabular}{l
  S[table-format = 3.2 \pm 2.2]
  S[table-format = 3.2 \pm 2.2]
  S[table-format = 3.2 \pm 2.2]
  S[table-format = 3.2 \pm 2.2]
  S[table-format = 3.2 \pm 2.2]}
\toprule
{Agents} & {0.2} & {0.5} & {0.8} & {1.0} & {2.0} \\
\midrule
5  & \num{4.72(0.35)} & \num{12.02(1.21)} & \num{19.99(1.53)} & \num{26.13(1.15)} & \num{65.56(2.96)} \\
10 & \num{5.91(0.11)} & \num{22.18(4.63)} & \num{37.82(0.92)} & \num{42.96(1.05)} & \num{68.04(1.09)} \\
15 & \num{9.15(0.95)} & \num{41.49(5.03)} & \num{56.66(4.15)} & \num{63.08(2.88)} & \num{99.46(4.82)} \\
20 & \num{10.07(2.06)}& \num{46.50(5.49)} & \num{57.06(6.07)} & \num{63.55(5.41)} & \num{103.71(12.46)} \\
\bottomrule
\end{tabular}
\caption{Mean and standard deviation of reported times (in seconds) of the \mbox{\Multihit} algorithm with $\mathrm{Random}$ parameter initialization on the benchmarks for each combination of number of agents (rows) and $r$ (columns).
The time complexity increases with both the number of agents and the number of evaluation steps.}
\label{tab:gd_times}
\end{table*}

\begin{table*}[tb]
\centering
\begin{tabular}{l
  S[table-format = 3.2 \pm 2.2]
  S[table-format = 3.2 \pm 2.2]
  S[table-format = 3.2 \pm 2.2]
  S[table-format = 3.2 \pm 2.2]
  S[table-format = 3.2 \pm 2.2]}

\toprule
{Agents} & {0.2} & {0.5} & {0.8} & {1.0} & {2.0} \\
\midrule
5  & \num{5.42(0.38)}  & \num{14.28(2.02)} & \num{27.43(3.31)} & \num{34.82(4.05)} & \num{63.34(7.57)} \\
10 & \num{7.31(0.22)}  & \num{24.81(3.00)} & \num{33.96(2.71)} & \num{38.50(2.93)} & \num{61.88(2.69)} \\
15 & \num{11.37(1.24)} & \num{40.72(4.61)} & \num{50.16(5.88)} & \num{56.65(5.66)} & \num{89.84(5.54)} \\
20 & \num{12.61(1.53)} & \num{41.86(5.74)} & \num{51.15(6.26)} & \num{56.97(6.39)} & \num{93.65(9.08)} \\
\bottomrule
\end{tabular}
\caption{Mean and standard deviation of reported times (in seconds) of the \mbox{\Multihit} algorithm with $\mathrm{LP}$ parameter initialization on the benchmarks for each combination of number of agents (rows) and $r$ (columns).
The time complexity increases with both the number of agents and the number of evaluation steps.}
\label{tab:lpgd_times}
\end{table*}

\begin{table*}[tb]
\centering
\begin{tabular}{l
  S[table-format = 3.2 \pm 2.2]
  S[table-format = 3.2 \pm 2.2]
  S[table-format = 3.2 \pm 2.2]
  S[table-format = 3.2 \pm 2.2]
  S[table-format = 3.2 \pm 2.2]}
\toprule
{Agents} & {0.2} & {0.5} & {0.8} & {1.0} & {2.0} \\
\midrule
5  & \num{25.22(3.05)}  & \num{0.99(0.04)}  & \num{0.99(0.04)}  & \num{0.99(0.04)}  & \num{1.00(0.04)} \\
10  & \num{22.82(2.70)}  & \num{0.98(0.04)}  & \num{0.97(0.04)}  & \num{0.98(0.04)}  & \num{0.99(0.04)} \\
15  & \num{19.87(2.91)}  & \num{0.97(0.04)}  & \num{0.97(0.04)}  & \num{0.97(0.04)}  & \num{0.98(0.04)} \\
20  & \num{18.74(2.81)}  & \num{0.96(0.05)}  & \num{0.96(0.04)}  & \num{0.96(0.04)}  & \num{0.97(0.04)} \\
\bottomrule
\end{tabular}
\caption{Mean and standard deviation of normalized values (value/base) of the \mbox{\Multihit} algorithm with $\mathrm{Random}$ parameter initialization on the benchmarks for each combination of number of agents (rows) and $r$ (columns). Lower values indicate better performance relative to the baseline.}
\label{tab:gd_values}
\end{table*}

\begin{table*}[tb]
\centering
\begin{tabular}{l
  S[table-format = 3.2 \pm 2.2]
  S[table-format = 3.2 \pm 2.2]
  S[table-format = 3.2 \pm 2.2]
  S[table-format = 3.2 \pm 2.2]
  S[table-format = 3.2 \pm 2.2]}
\toprule
{Agents} & {0.2} & {0.5} & {0.8} & {1.0} & {2.0} \\
\midrule
5  & \num{1.00(0.00)}  & \num{0.97(0.03)}  & \num{0.97(0.03)}  & \num{0.97(0.03)}  & \num{0.97(0.03)} \\
10  & \num{1.00(0.00)}  & \num{0.96(0.04)}  & \num{0.96(0.04)}  & \num{0.96(0.04)}  & \num{0.96(0.04)} \\
15  & \num{1.00(0.00)}  & \num{0.96(0.05)}  & \num{0.96(0.05)}  & \num{0.96(0.05)}  & \num{0.96(0.05)} \\
20  & \num{1.00(0.00)}  & \num{0.95(0.05)}  & \num{0.95(0.05)}  & \num{0.95(0.05)}  & \num{0.95(0.05)} \\
\bottomrule
\end{tabular}
\caption{Mean and standard deviation of normalized values (value/base) of the \mbox{\Multihit} algorithm with $\mathrm{LP}$ parameter initialization on the benchmarks for each combination of number of agents (rows) and $r$ (columns). Lower values indicate better performance relative to the baseline.}
\label{tab:lp_gd_values}
\end{table*}

\paragraph{Results.}
In Table~\ref{tab:gd_times} we report the mean and standard deviation of the total optimization time of the \mbox{\Multihit}
algorithm with $\mathrm{Random}$ parameter initialization.
The statistics are computed separately for each number of agents and each value of $r$, aggregated over all benchmarks and all runs.
In Table~\ref{tab:lpgd_times} we report the same statistics for $\mathrm{LP}$ initialization.

In Table~\ref{tab:gd_values} we report the same statistics, but of the values of the strategy profiles found, normalized by the value of the baseline, i.e., the values are $\Val_{\RND} / \Base$.
In Table~\ref{tab:lp_gd_values}  we report the same statistics for $\mathrm{LP}$ initialization.

\paragraph{Discussion.}
We can clearly see that the runtime increases with the number of agents and with $r$.
This is expected, as the number of terms in the sum of $\pmb{\Exp\langle\gamma\rangle_\pi^I[\MHit]}$ is precisely $\gamma_r$, and the number of factors in each product equals the number of agents.

Observe that for each $r$, the paths taken into account when computing $\pmb{\Exp\langle\gamma\rangle_\pi^I[\MHit]}$ have length at most $\gamma_r$.
From the tables on value, we see that for $\gamma_{0.2}$ the values obtained with random initialization are poor, while for LP initialization the values for $\gamma_{0.2}$ coincide with the LP baseline.
For our benchmarks, the value $\gamma_{0.2}$ coincides with the length of the shortest paths from the initial state to the goal.
Hence, only a small number of paths are taken into account when computing $\pmb{\Exp\langle\gamma\rangle_\pi^I[\MHit]}$, and the algorithm \mbox{\Multihit} cannot substantially improve the strategy value.

For larger values of $r$ that cover a larger set of paths, Table~\ref{tab:gd_values} and Table~\ref{tab:lp_gd_values} show that the influence of the actual choice of $r$ is negligible.
All values $\gamma_r$ with $r \geq 0.5$ yield identical strategy values for LP initialization.
For random initialization, the values are close to each other but appear to slightly decrease as $r$ increases.
A possible explanation is that larger values of $r$ incorporate longer paths into the objective, which increases the complexity of the optimization loop and may slow down convergence for a fixed optimization budget.

Thus, we conclude that $\gamma$ should be chosen large enough to cover a sufficiently rich set of paths.
As stated in the main text, we use $\gamma_{1.0}$, as it guarantees coverage of all simple paths.
Beyond this point, increasing $\gamma$ does not lead to a noticeable improvement in the performance of \mbox{\Multihit}.

\end{document}